\documentclass[pra,eqsecnum,twocolumn,amsmath,amssymb]{revtex4}

\usepackage{graphicx}
\usepackage{bm}
\usepackage[usenames,dvipsnames]{color}
\definecolor{altncolor}{rgb}{0,0,0.8}
\usepackage[colorlinks=true, linkcolor=green, anchorcolor=altncolor,
citecolor=altncolor, filecolor=altncolor, menucolor=altncolor,
urlcolor=altncolor]{hyperref}

\begin{document}


\author{Peter B. Weichman}

\affiliation{BAE Systems, Technology Solutions, 600 District
Avenue, Burlington, MA 01803}

\title{Strong vorticity fluctuations and antiferromagnetic correlations in axisymmetric fluid equilibria}

\date{\today}

\begin{abstract}

The macroscale structure and microscale fluctuation statistics of late-time asymptotic steady state flows in cylindrical geometries is studied using the methods of equilibrium statistical mechanics. The axisymmetric assumption permits an effective two-dimensional description in terms of the (toroidal) flow field $\sigma$ about the cylinder axis and the vorticity field $\xi$ that generates mixing within the (poloidal) planes of fixed azimuth. As for a number of other 2D fluid systems, extending the classic 2D Euler equation, the flow is constrained by an infinite number of conservation laws, beyond the usual kinetic energy and angular momentum. All must be accounted for in a consistent equilibrium description. It is shown that the most directly observable impact of the conservation laws is on $\sigma$, which displays interesting large-scale radius-dependent flow structure. However, unlike in some previous treatments, we find that the thermodynamic temperature is always positive. As a consequence, except for an infinitesimal boundary layer that maintains the correct (conserved) value of the overall poloidal circulation, the impact on $\xi$ resides in the statistics of the strongly fluctuating, fine-scale mixing, where it is sensitive to ``antiferromagnetic'' microscale correlations that help maintain the analogue of local charge neutrality. The poloidal flow is macroscopically featureless, displaying no large scale circulating jet- or eddy-like features (which typically emerge as negative temperature states in analogous Euler and quasigeostrophic equilibria).

\end{abstract}

%

\maketitle

\section{Introduction}
\label{sec:intro}

The modern era of exact statistical treatments of the late-time steady states of 2D fluid flows, properly accounting for the infinite number of conserved integrals of the motion, began with the Miller--Robert--Sommeria (MRS) theory of the 2D Euler equation \cite{M1990,RS1991,MWC1992,LB1967}, generalizing earlier approximate treatments going all the way back to the seminal work of Onsager \cite{O1949}, and progressing through the Kraichnan Energy--Enstrophy theory \cite{K1975}, various formulations of the point vortex problem (see, e.g., \cite{MJ1974,LP1977}), and extensions to the quasigeostrophic equations \cite{MR1994,BS2002,W2006}. Since then, the theory has been applied to significantly more complex systems, containing multiple interacting fields (in contrast to the Euler equation, which reduces to a single scalar equation for the vorticity), but still possessing an infinite number of conserved integrals \cite{HMRW1985}. These include magnetohydrodynamic equilibria \cite{JT1997,W2012}, the shallow water equations \cite{WP2001,CS2002,W2017}, as well as numerous other geophysical applications \cite{BV2012}.

\begin{figure}

\includegraphics[width=3.2in,viewport=190 90 815 490,clip]{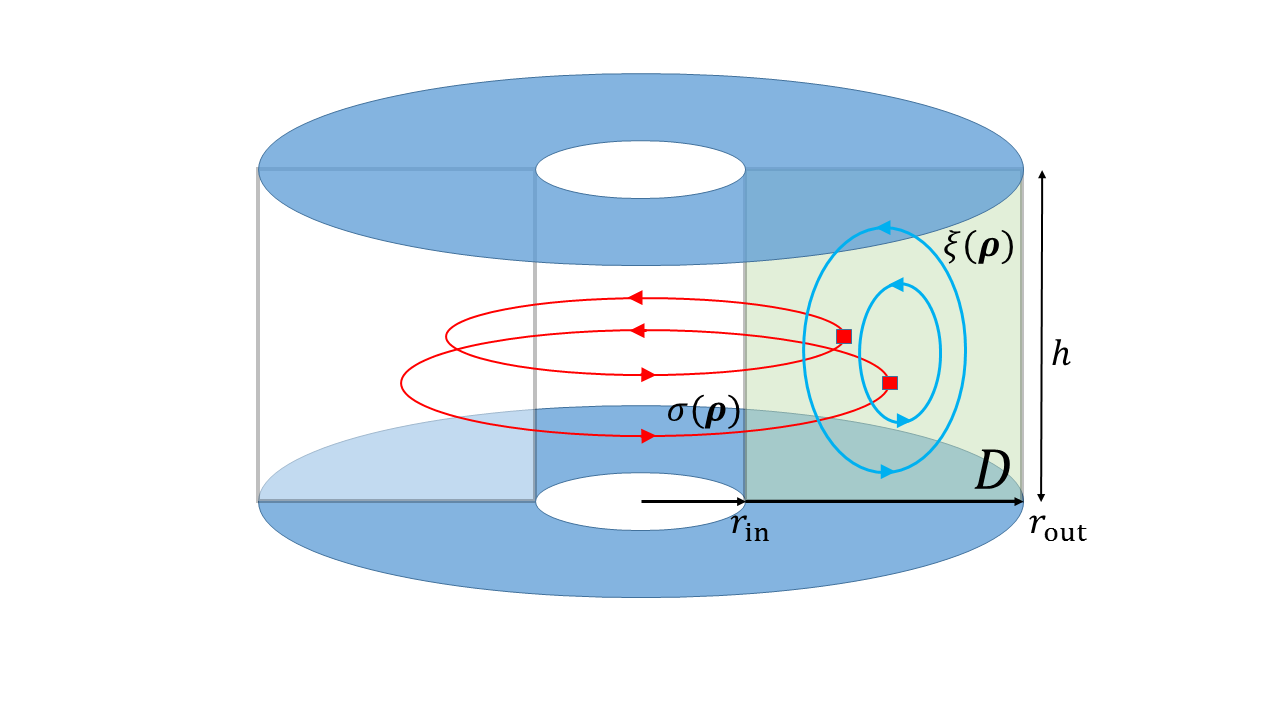}

\caption{Axisymmetric flow geometry. The pattern of flows is invariant under rotation about the cylinder axis, and hence may be fully specified by the toroidal flow field $\sigma$ about the axis [equation (\ref{2.7})], and poloidal vorticity field $\xi$ [equation (\ref{2.9})] within any 2D radial plane $D$.}

\label{fig:axisymgeom}
\end{figure}

Here we return to the example of 3D axisymmetric flows, refining and substantially extending a theory presented in \cite{TDB2014}. These flows, sketched in Fig.\ \ref{fig:axisymgeom}, are interesting because they allow constrained interaction between circulation about the cylinder axis (toroidal flow), and circulation within any given fixed 2D azimuth plane containing that axis (poloidal flow). A novel aspect of this system, making it substantially different from the classic case of the Euler equation, is that although the key 2D dynamics occur in the ``poloidal plane'', the vorticity field $\xi$ in that plane is not conserved by the flow. Rather, it is the toroidal field $\sigma$ that is passively advected, and is then constrained by an infinite number of conserved integrals. It is therefore the indirect effects of $\sigma$-conservation that must feed back on $\xi$ to generate any interesting equilibrium poloidal flow structure. One consequence is that, due to the comparatively weak constraints on $\xi$, the equilibrium state is strongly fluctuating and only positive temperature equilibria are permitted. This is in contrast to the Euler case in which the direct constraints on the vorticity field strongly suppress fluctuations, the equilibria may derived from an exact variational (mean field) theory, and both positive and negative temperature equilibria are permitted (with the latter sometimes leading to large-scale vortex structures). It will be seen that in the axisymmetric flow case, the variational approximation is not generally applicable.

\subsection{Outline and summary of results}
\label{sec:outline}

The outline of the remainder of this paper is as follows. The basic equations of motion, derived from the 3D Euler equation under the axial flow assumption and reduced to a pair of coupled 2D equations for $\xi$ and $\sigma$ (defined on the poloidal plane), are summarized in Sec.\ \ref{sec:bkgnd}. The key conservation laws are summarized in Sec.\ \ref{sec:conserve}. In addition to the usual kinetic energy, linear momentum along the cylinder axis, and angular momentum about the axis, these include two infinite classes of conserved circulation integrals that are a consequence of the constrained effectively 2D dynamics. The first class involves $\sigma$ alone. The second couples both fields, but only linearly in $\xi$. All of the conserved values may be viewed as fixed by the flow initial condition, and then maintained through the turbulent cascade that leads to the late-time equilibrated flow.

In Sec.\ \ref{sec:eqmfe} the thermodynamic free energy fully characterizing the equilibrium flows is defined in terms of the underlying grand canonical statistical mechanics formalism laid out in Apps.\ \ref{app:liouville} and \ref{app:statmech}. The continuum fluid results are obtained through a limiting procedure in which the continuous poloidal plane is replaced by a 2D grid with microscopic mesh size $a$, and the limit $a \to 0$ taken. Similar to the Euler case \cite{MWC1992}, it transpires that the temperature and other model parameters must be scaled appropriately with $a$ in order to obtain a sensible thermodynamic limit. At this point the basic positive temperature requirement is an obvious consequence of the unboundedness of the poloidal field $\xi$, which is then capable of absorbing arbitrarily energies.

To gain basic intuition, in Sec.\ \ref{sec:decoupledmodel} the exact solution is derived for the special choice of parameters in which the toroidal and poloidal degrees of freedom are completely decoupled. This solution will turn out to be highly relevant to the general case as well. We also introduce a class of ``finite level'' models in which the full set of conserved integrals are constrained to a finite number through a special choice of parameters. These are particularly amenable to analytic and numerical solution.

In Sec.\ \ref{sec:genmodel} the full coupled model is considered, and reduced models are derived focusing separately on the toroidal and poloidal fields. In particular, by integrating out the poloidal field $\xi$, the effective theory for $\sigma$ is shown to be equivalent to a certain type of classical antiferromagnetic (AF) spin model on the microgrid. Proper scaling in the $a \to 0$ limit, however, turns out to generate only very weak antiferromagnetic correlations, allowing an exact perturbative treatment of both classes of conservation laws. The intuition here is that the bias on $\xi$ introduced by the second class of conservation laws is rather weak compared to the intrinsic microscale fluctuations, and the final result lies very close to the decoupled model solution.

On the other hand, the effective theory for $\xi$, obtained by integrating out the toroidal field $\sigma$, is that of a plasma model with long range logarithmic (2D Coulomb-type) interactions---as would be expected from the underlying vortex degrees of freedom. The AF correlations are here reflected in the strong tendency toward local charge neutrality in Coulomb systems, which includes also the classic electrostatic effect in which all uncompensated free charges are pushed to the system boundary. There they are distributed in such a way as to produce vanishing interior large scale flow (equivalent to the requirement of vanishing static electric field within a conducting body).

In Sec.\ \ref{sec:apps} some simple examples, including a detailed look at the two-level model, are treated and used to illustrate the general theory.

The paper is concluded in Sec.\ \ref{sec:conclude}. In particular, comparison is made with the approach of Ref.\ \cite{TDB2014} in which an artificial cutoff $|\xi| < M$ is applied, and the limit $M \to \infty$ taken after $a \to 0$. These two limits do not generally commute and this strongly impacts the physical consequences of the theory. In particular, the variational approach used there, indeed valid in parameter regimes where $|\xi| = O(M)$ is dominated by the cutoff, is seen here to fail when $|\xi| \ll M$ is finite, missing the effects of strong positive temperature fluctuations.

\section{Background}
\label{sec:bkgnd}

Axisymmetric flows are confined to a 3D domain of revolution with cylindrical coordinates $(\theta,r,z) \in {\cal D} \equiv [0,2\pi) \times D$, where $D$ is a fixed 2D domain in the $rz$-plane (Fig.\ \ref{fig:axisymgeom}). Ultimately, we will specialize to a finite cylindrical domain, $R_\mathrm{in} \leq r \leq R_\mathrm{out}$, with periodic boundary conditions along the axis of the cylinder $0 \leq z \leq h$, but for now $D$ is arbitrary.

The flows obey the three-dimensional Euler equation
\begin{equation}
\partial_t {\bf v} + ({\bf v} \cdot \nabla) {\bf v} = -\nabla p
\label{2.1}
\end{equation}
in which the pressure is uniquely determined by the incompressibility constraint
\begin{equation}
\nabla \cdot {\bf v} = 0,
\label{2.2}
\end{equation}
but specialized to velocity fields
\begin{equation}
{\bf v} = v_r(r,z) \hat {\bf r}
+ v_z(r,z) \hat {\bf z}
+ v_\theta(r,z) \hat {\bm \theta}
\label{2.3}
\end{equation}
whose cylindrical coordinate components are independent of the azimuthal angle $\theta$.

\subsection{Vorticity and stream function}
\label{sec:vortstream}

It follows from (\ref{2.2}) that
\begin{equation}
\partial_r (r v_r) + \partial_z (r v_z) = 0,
\label{2.4}
\end{equation}
allowing one to represent
\begin{equation}
v_r = -\frac{1}{r} \partial_z \psi,\ \
v_z = \frac{1}{r} \partial_r \psi
\label{2.5}
\end{equation}
in terms of a stream function $\psi(r,z)$. The velocity is therefore conveniently represented in the form
\begin{equation}
{\bf v} = \nabla \times \left(\frac{1}{r} \psi \hat {\bf \theta} \right)
+ \frac{1}{r} \sigma \hat {\bm \theta}
\label{2.6}
\end{equation}
in which $\psi$ and the vertical component of the angular momentum density
\begin{equation}
\sigma = r v_\theta
\label{2.7}
\end{equation}
about the symmetry axis are taken as the fundamental fields.

The axial vorticity $\omega_\theta$ is related to the stream function by
\begin{eqnarray}
\omega_\theta &\equiv& \hat {\bf z} \cdot \nabla \times {\bf v}
= \partial_z v_r - \partial_r v_z
\nonumber \\
&=& - \frac{1}{r} \partial_z^2 \psi
- \partial_r \left(\frac{1}{r} \partial_r \psi \right).
\label{2.8}
\end{eqnarray}
Defining scaled vorticity
\begin{equation}
\xi = \frac{\omega_\theta}{r},
\label{2.9}
\end{equation}
and the modified radial and 2D coordinates
\begin{equation}
y = \frac{r^2}{2},\ {\bm \rho} = (y,z)
\label{2.10}
\end{equation}
one may express the poloidal velocity components in the form
\begin{eqnarray}
(v_r,v_z) &=& \left(-\frac{1}{\sqrt{2y}} \partial_z \psi,  \partial_y \psi \right)
\equiv \nabla_* \times \psi
\nonumber \\
\nabla_* &\equiv& \hat {\bf y} \partial_y \psi
+ \hat {\bf z} \frac{1}{\sqrt{2y}} \partial_z,
\label{2.11}
\end{eqnarray}
and the poloidal (scaled) vorticity is given by
\begin{equation}
\xi = -\left(\frac{1}{2y} \partial_z^2 + \partial_y^2 \right) \psi
\equiv - \Delta_* \psi.
\label{2.12}
\end{equation}
For a simply connected domain, the free slip condition is enforced with Dirichlet boundary conditions, $\psi|_{\partial D} = 0$. The generalized Laplace equation may be solved to derive $\psi$ from any given $\xi$:
\begin{equation}
\psi({\bm \rho}) = \int_D d{\bm \rho}'
G({\bm \rho},{\bm \rho}') \xi({\bm \rho})
\label{2.13}
\end{equation}
with generalized Laplacian Green function obeying
\begin{equation}
-\Delta_* G({\bm \rho},{\bm \rho}')
= \delta({\bm \rho} - {\bm \rho}').
\label{2.14}
\end{equation}
and satisfying the same Dirichlet boundary condition in both arguments.

More generally, for the multiply connected cylindrical geometry of primary interest here, free slip boundary conditions in general allow different constant values of $\psi$ on the inner and outer boundaries of the cylinder, $r_\mathrm{in} \leq r \leq r_\mathrm{out}$ \cite{W2017}. This may be handled by accounting for the mean vertical flow
\begin{equation}
v_z^0 = \frac{1}{V_{\cal D}} \int_{\cal D} d{\bf x} v_z
\label{2.15}
\end{equation}
in which $V_{\cal D} = \pi h(r_\mathrm{out}^2 - r_\mathrm{in}^2) = 2\pi h(y_\mathrm{out} - y_\mathrm{in})$ is the cylinder volume, with $h$ its height. Conservation of vertical momentum implies that $v_z^0$ is a constant of the motion (see Sec.\ \ref{sec:zmomentum}), and the subtracted stream function
\begin{equation}
\psi_D = \psi - v_z^0 y,\
\label{2.16}
\end{equation}
may be chosen to vanish on both boundaries. The scaled vorticity remains unchanged,
\begin{equation}
{\bf v}_D = {\bf v} - v_z^0 \hat {\bf z}\
\Rightarrow\ \xi = -\Delta_* \psi_D
\label{2.17}
\end{equation}
and (\ref{2.13}) therefore generalizes to
\begin{eqnarray}
\psi({\bm \rho}) &=& \psi_D({\bm \rho}) + v_z^0 y
\nonumber \\
&=& \int_D d{\bm \rho}'
G({\bm \rho},{\bm \rho}') \xi({\bm \rho}) + v_z^0 y
\label{2.18}
\end{eqnarray}
in which $v_z^0$ is a fixed parameter defined, e.g., by the flow initial condition.

\subsection{Equations of motion in terms of $\xi$ and $\sigma$}
\label{sec:eqmotxisigma}

The azimuthal component of the Euler equation takes the form
\begin{equation}
\partial_t v_\theta + v_r \partial_r v_\theta
+ v_z \partial_z v_\theta + \frac{v_\theta v_r}{r} = 0,
\label{2.19}
\end{equation}
which reduces to
\begin{equation}
\partial_t \sigma + {\bf w} \cdot \nabla_{\bm \rho} \sigma = 0,
\label{2.20}
\end{equation}
in which $\nabla_{\bm \rho} = (\partial_y,\partial_z)$ is a 2D gradient, and
\begin{equation}
{\bf w} = \nabla_{\bm \rho} \times \psi = (-\partial_z \psi, \partial_y \psi)
= (rv_r,v_z)
\label{2.21}
\end{equation}
satisfies the 2D incompressibility condition
\begin{equation}
\nabla_{\bm \rho} \cdot {\bf w} = 0.
\label{2.22}
\end{equation}
Equation (\ref{2.20}) expresses the fact that, in this modified 2D coordinate system, the toroidal flow parameter $\sigma$ is freely advected by the incompressible poloidal flow generated by $\xi$.

Applying appropriate spatial derivatives to the Euler equation and using the incompressibility constraint (\ref{2.4}), the axial vorticity (\ref{2.8}) obeys
\begin{equation}
\partial_t \omega_\theta + v_r \partial_r \omega_\theta
+ v_z \partial_z \omega_\theta - \frac{\omega_\theta v_r}{r}
= \frac{\partial_z v_\theta^2}{r},
\label{2.23}
\end{equation}
which may be put in the form
\begin{equation}
\partial_t \xi + {\bf w} \cdot \nabla_{\bm \rho} \xi
= \frac{\partial_z \sigma^2}{4y^2}.
\label{2.24}
\end{equation}
Thus, 2D advection of $\xi$ (by its self-generated poloidal flow field) is forced by $\sigma$.

Equations (\ref{2.20}) and (\ref{2.24}) are the fundamental equations of motion for axisymmetric flow, reducing the 3D vector Euler equation (\ref{2.1}) to a pair of coupled 2D scalar equations. The original 3D velocity field is recovered using (\ref{2.7}), and by constructing the 2D stream function (\ref{2.18}), and then using (\ref{2.5}) in conjunction with the coordinate mapping (\ref{2.10}).

\section{Conservation laws}
\label{sec:conserve}

\subsection{Conserved energy}
\label{sec:cons_KE}

The kinetic energy is
\begin{eqnarray}
E &=& \frac{1}{2} \int_{\cal D} d{\bf x} |{\bf v}|^2
= \frac{1}{2} \int_{\cal D} d{\bf x} 
\left[|{\bf v}_D|^2 + (v_z^0)^2 \right]
\nonumber \\
&=& \pi \int_D d{\bm \rho} \left[|\nabla_* \psi_D|^2
+ \frac{\sigma^2}{2y} \right] + E_z^0
\nonumber \\
&=& \pi \int d{\bm \rho} \left(\xi \psi_D + \frac{\sigma^2}{2y} \right) + E_z^0
\nonumber \\
&=& E_G[\xi] + \pi \int d{\bm \rho} \frac{\sigma^2}{2y} + E_z^0
\label{3.1}
\end{eqnarray}
in which
\begin{eqnarray}
E_z^0 &=& \frac{1}{2} V_{\cal D} (v_z^0)^2
\nonumber \\
E_G[\xi] &=& \pi \int_D d{\bm \rho} \int_D d{\bm \rho}'
\xi({\bm \rho}) G({\bm \rho},{\bm \rho}') \xi({\bm \rho}')
\label{3.2}
\end{eqnarray}
are the kinetic energies associated, respectively, with the mean and poloidal flows. To derive the latter, equation (\ref{2.18}) has been used following an integration by parts. The surface term vanishes by virtue of the ability to impose $\psi_D|_{\partial D} = 0$.

Energy conservation (which does not require axisymmetry) is verified by noting that, using the Euler equation (\ref{2.1}) and the incompressibility condition (\ref{2.2}), the equation of motion for the kinetic energy density
\begin{equation}
\varepsilon = \frac{1}{2}|{\bf v}|^2
\label{3.3}
\end{equation}
takes the form of the conservation law
\begin{equation}
\partial_t \varepsilon + \nabla \cdot {\bf j}_\varepsilon = 0
\label{3.4}
\end{equation}
with energy current
\begin{equation}
{\bf j}_\varepsilon = (p + \varepsilon) {\bf v}.
\label{3.5}
\end{equation}
It follows that
\begin{eqnarray}
\partial_t E &=& -\int_{\cal D} d{\bf x}
\nabla \cdot {\bf j}_\varepsilon
\nonumber \\
&=& -\int_{\partial {\cal D}} dA
(p + \varepsilon) {\bf v} \cdot \hat {\bf n} = 0.
\label{3.6}
\end{eqnarray}
The surface integral over the boundary $\partial {\cal D}$ vanishes for any combination of periodic and free slip boundary conditions (${\bf v} \cdot \hat {\bf n} = 0$).

\begin{figure}

\includegraphics[width=3.2in,viewport=190 50 780 490,clip]{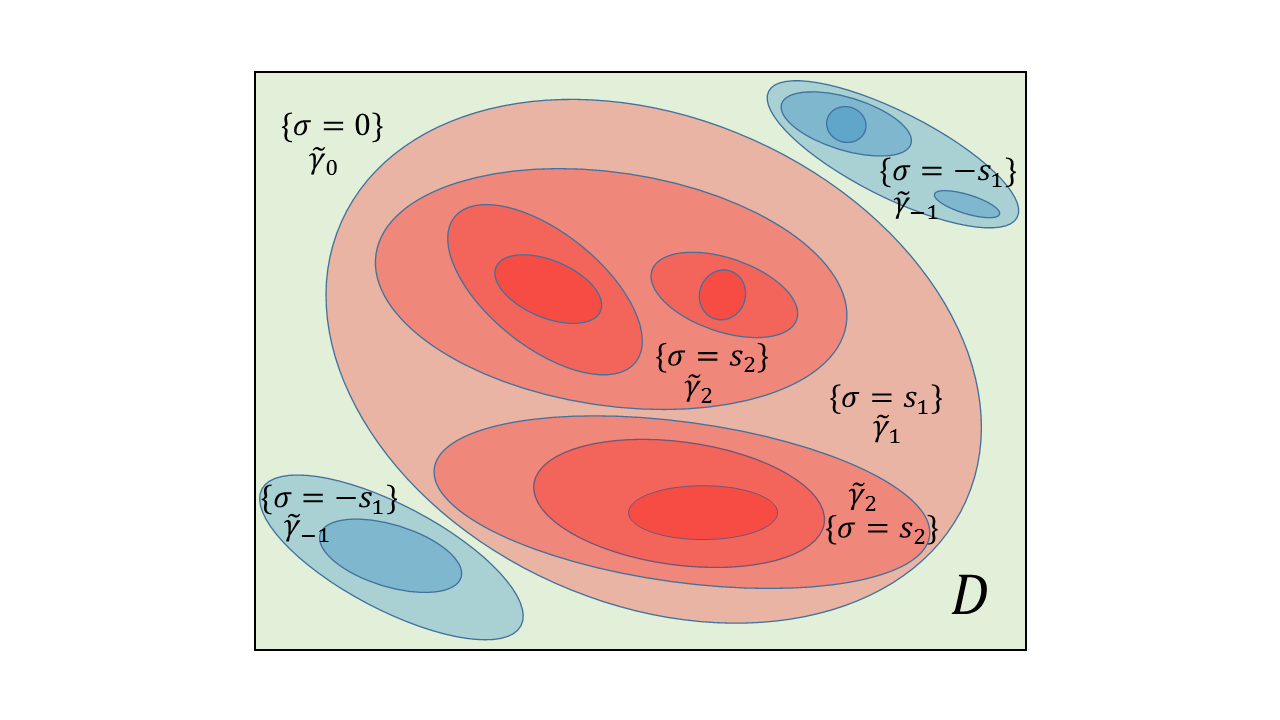}

\caption{Illustration of conserved vorticity integrals. As formally defined by (\ref{3.15}), the total area $\gamma[\sigma](s)$ of each level set $\{\sigma({\bm \rho}) = s\}$ is conserved by the flow, as is the total integral $\tilde \gamma[\sigma,\xi](s)$ of $\xi$ over each such set. In general each set is the union of some number of (in general, multiply connected) pieces. For clarity, only a few of these are explicitly labeled in the figure. Since the exact value of $\sigma$, but only the mean value of $\xi$, is specified on each set, the former is far more strongly constrained. Arbitrary fluctuations of the latter about the mean are permitted, and this greatly impacts the predicted equilibrium states.}

\label{fig:conserve}
\end{figure}

\subsection{Conserved vorticity integrals}
\label{sec:cons_vorticity}

It follows from (\ref{2.20}) and (\ref{2.22}) that any function $f(\sigma)$ obeys the conservation law
\begin{equation}
\partial_t f(\sigma) + \nabla_{\bm \rho} \cdot [f(\sigma) {\bf w}] = 0
\label{3.7}
\end{equation}
The vorticity integral
\begin{equation}
\Omega_f = \pi \int_D d{\bm \rho} f(\sigma)
\label{3.8}
\end{equation}
then obeys
\begin{eqnarray}
\partial_t \Omega_f &=& -\pi \int_D d{\bm \rho}
\nabla_{\bm \rho} \cdot [f(\sigma) {\bf w}]
\nonumber \\
&=& -\pi \int_{\partial D} dl f(\sigma) {\bf w} \cdot \hat {\bf n} = 0
\label{3.9}
\end{eqnarray}
Here, we make use of the fact that the free slip boundary condition is equivalent to ${\bf w} \cdot \hat {\bf n} = 0$ in the modified coordinates, and is valid for arbitrary $D$. The special case $f(\sigma) = \sigma$ coincides with conservation of total angular momentum about the symmetry axis.

More generally, from  (\ref{2.20}), (\ref{2.22}) and (\ref{2.24}) the combination $\xi f(\sigma)$ obeys
\begin{equation}
\partial_t [\xi f(\sigma)]
+ \nabla_{\bm \rho} \cdot [\xi f(\sigma) {\bf w}]
= \frac{f(\sigma) \partial_z \sigma^2}{4y^2}
\label{3.10}
\end{equation}
from which follows the conservation law form
\begin{equation}
\partial_t [\xi f(\sigma)]
+ \nabla_{\bm \rho} \cdot \left[\xi f(\sigma) {\bf w}
+ \frac{F(\sigma)}{4y^2} \hat {\bf z} \right] = 0
\label{3.11}
\end{equation}
in which $F(\sigma)$ obeys
\begin{equation}
F'(\sigma) = 2\sigma f(\sigma) \ \Rightarrow \
F(\sigma) = 2 \int_0^\sigma s f(s) ds.
\label{3.12}
\end{equation}

The integral
\begin{equation}
\tilde \Omega_f = \pi \int_D d{\bm \rho} \xi f(\sigma)
\label{3.13}
\end{equation}
therefore obeys
\begin{eqnarray}
\partial_t \tilde \Omega_f
&=& -\pi \int_{\partial D} dl \frac{F(\sigma)}{4y^2}
\hat {\bf z} \cdot \hat {\bf n}
\label{3.14}
\end{eqnarray}
which vanishes for a cylindrical boundary $r_\mathrm{in} \leq r \leq r_\mathrm{out}$ (enforcing $\hat {\bf z} \cdot \hat {\bf n} = 0$) along with periodic boundary conditions in $0 \leq z < h$.

The interpretation here is that in addition to being self-advected by ${\bf w}$, the axial vorticity $\xi$ is advected vertically by the angular momentum density $\sigma$. Hence, only for a vertical boundary does the net advection keep $\xi$ from effectively exiting or entering the domain.

As illustrated in Fig.\ \ref{fig:conserve}, the full infinite set of conserved vorticity integrals may be conveniently parameterized by the functionals
\begin{eqnarray}
\gamma[\sigma](s) &=& \int_D d{\bm \rho} \delta[s - \sigma({\bm \rho})]
\nonumber \\
\tilde \gamma[\xi,\sigma](s) &=& \int_D d{\bm \rho} \xi({\bm \rho})
\delta[s - \sigma({\bm \rho})]
\label{3.15}
\end{eqnarray}
which measure, for each real value $-\infty < s < \infty$, the fractional area on which $\sigma$ takes the value $s$, in the second case weighted by the values of $\xi$. Given these functions, one directly derives for any function $f$,
\begin{eqnarray}
\Omega_f[\sigma] &=& \int ds \gamma[\sigma](s) f(s)
\nonumber \\
\tilde \Omega_f[\xi,\sigma] &=& \int ds \tilde \gamma[\xi,\sigma](s) f(s).
\label{3.16}
\end{eqnarray}

\subsection{Conserved vertical momentum}
\label{sec:zmomentum}

As alluded to below (\ref{2.15}), for the case of cylindrical boundary, translation symmetry in $z$ implies conservation of vertical momentum
\begin{equation}
P_z = \int_{\cal D} d{\bf x} v_z = V_{\cal D} v_z^0,
\label{3.17}
\end{equation}
with or without axisymmetry. To verify this formally, the $z$-component of (\ref{2.1}) may be written in the form
\begin{equation}
\partial_t v_z + \nabla \cdot {\bf j}_z = 0,\ \
{\bf j}_z \equiv v_z {\bf v} + p \hat {\bf z},
\label{3.18}
\end{equation}
and it follows that
\begin{equation}
\partial_t P_z = -\int_{\partial D} dA
(v_z {\bf v} + p \hat {\bf z}) \cdot \hat {\bf n} = 0
\label{3.19}
\end{equation}
for free slip boundary conditions on a cylinder.

From (\ref{2.5}) it follows more explicitly that
\begin{equation}
P_z = 2\pi h [\psi(r_\mathrm{out}) - \psi(r_\mathrm{in})]
\label{3.20}
\end{equation}
is completely determined by the Dirichlet boundary conditions on $\psi$. In particular, as claimed earlier, $P_z = 0$ may be enforced by setting $\psi = 0$ on both boundaries.

\section{Equilibrium free energy}
\label{sec:eqmfe}

The grand canonical partition function, defined by (\ref{B1})--(\ref{B3}), takes the form
\begin{eqnarray}
Z &=& e^{-\beta E_z^0} Z_0[\beta,\mu,\tilde \mu; h_\sigma,h_\xi]
\nonumber \\
Z_0 &\equiv& \int D[\sigma] e^{-\beta \int_D d{\bm \rho}
\left[\frac{\pi \sigma^2}{2y} - \mu(\sigma)
- h_\sigma({\bm \rho}) \sigma \right]}
\nonumber \\
&&\times\ \int D[\xi] e^{-\beta \left\{E_G[\xi]
- \int_D d{\bm \rho} \xi [\tilde \mu(\sigma) + h_\xi({\bm \rho})] \right\}},
\ \ \ \ \
\label{4.1}
\end{eqnarray}
with corresponding free energy
\begin{eqnarray}
F &=& E_z^0 + F_0[\beta,\mu,\tilde \mu; h_\sigma,h_\xi]
\nonumber \\
F_0 &\equiv& -\frac{1}{\beta} \ln(Z_0),
\label{4.2}
\end{eqnarray}
in which the functional integrals are defined by the limit (\ref{A2}), and $E_z^0, E_G$ were defined in (\ref{3.2}). The conserved quantities are obtained from the free energy derivatives (\ref{B5}) with respect to the Lagrange multipliers $\beta,\mu,\tilde \mu$. We have also included conjugate fields $h_\sigma$ and $h_\xi$ that will be set to zero in the end, but whose free energy derivatives may be used to generate statistical averages and correlations of the two fields.

We will consider first, in Sec.\ \ref{sec:decoupledmodel}, the model $\tilde \mu \equiv 0$, in which the two fields are entirely decoupled. This model is exact in non-cylindrical geometries, where the second class of conserved integrals is absent and $\tilde \mu$ therefore does not appear. This limit also serves to illustrate the nature of the limit $a \to 0$. We will then consider in Sec.\ \ref{sec:genmodel} the full coupled model, in particular deriving reduced models by first integrating out either one of $\xi$ or $\sigma$. The statistics of the remaining field exhibit rather different physical phenomena.

\section{Decoupled model}
\label{sec:decoupledmodel}

Setting $\tilde \mu \equiv 0$, as well as $h_\sigma = h_\xi = 0$ for now, one obtains
\begin{equation}
F_0(\beta,\mu) = F_\sigma(\beta,\mu) + F_\xi(\beta)
\label{5.1}
\end{equation}
in which
\begin{eqnarray}
F_\sigma &=& -\frac{1}{\beta} \ln(Z_\sigma)
\nonumber \\
F_\xi &=& -\frac{1}{\beta} \ln(Z_\xi)
\label{5.2}
\end{eqnarray}
with decoupled partition functions
\begin{eqnarray}
Z_\sigma &=& \int D[\sigma] e^{-\beta \int_D d{\bm \rho}
\left[\frac{\pi \sigma^2}{2y} - \mu(\sigma) \right]}
\nonumber \\
Z_\xi &=& \int D[\xi] e^{-\beta E_G[\xi]}.
\label{5.3}
\end{eqnarray}
We analyze each in sequence, with special attention to the continuum limit, $a \to 0$. Note that when considered separately the two models may be well defined over different temperature ranges (e.g., only positive vs.\ either sign), but in the end the sum (\ref{5.1}) constrains the thermodynamics to a common temperature range for which both are defined (e.g., only positive).

\subsection{Statistics of $\sigma$}
\label{sec:statsig}

Using the square grid discretization (\ref{A2}) \cite{foot:gridunits}, the statistics of the $\sigma$ field are clearly decoupled from site to site, and one obtains
\begin{equation}
Z_\sigma = \prod_l Z_1[\beta a^2,\mu,y_l]
\label{5.4}
\end{equation}
with factor $a^2$ coming from the discretization of $\int_D d{\bm \rho}$, the product extending over all lattice sites, and single site partition function
\begin{equation}
Z_1(\bar \beta,\mu,y) = \int ds e^{-\bar \beta
\left[\frac{\pi s^2}{2y} - \mu(s) \right]}.
\label{5.5}
\end{equation}
Typically, the flow initial condition will have bounded $\sigma$, hence $\gamma[\sigma](s)$ has compact support, and so therefore will $e^{-\bar \beta \mu(s)}$. Thus, (\ref{5.5}) is strongly convergent for any real $\bar \beta$, both positive and negative. The single site probability distribution takes the form
\begin{equation}
p_\sigma(s,y) = \langle \delta[s - \sigma({\bm \rho})] \rangle
= \frac{e^{-\beta a^2
\left[\frac{\pi s^2}{2y} - \mu(s) \right]}}{Z_1(\beta a^2,\mu,y)},
\label{5.6}
\end{equation}
and the position-dependent mean is
\begin{equation}
\langle \sigma({\bm \rho}) \rangle = \int s ds p_\sigma(s,y).
\label{5.7}
\end{equation}

Equation (\ref{5.2}) serves to illustrate another key result. It is seen that the \emph{properly scaled thermal variable} is
\begin{equation}
\bar \beta = \beta a^2\ \Rightarrow\ \bar T = T/a^2,
\label{5.8}
\end{equation}
which needs to remain finite in the continuum limit $a \to 0$ \cite{MWC1992}. Only for finite $\bar \beta$ does $\sigma$ have a nontrivial distribution. The $\sigma$ field free energy
\begin{eqnarray}
F_\sigma(\bar \beta,\mu) &=& -\lim_{a \to 0} \frac{1}{\beta} \ln(Z_\sigma)
\nonumber \\
&=& -\frac{1}{\bar \beta}
\int_D d{\bm \rho} \ln[Z_1(\bar \beta,\mu,y)] \ \ \ \ \
\label{5.9}
\end{eqnarray}
is also finite in the continuum limit. The latter corresponds to the thermodynamic limit in the sense that the number of grid cells $V_{\cal D}/\pi a^2$ diverges. For cylindrical ${\cal D}$ this expression may clearly be simplified, but in this section we treat this as a special case of the more general domain.

Finite $\beta$, on the other hand, leads to $\bar \beta = \beta/a^2 \to \infty$ and $\bar T = Ta^2 \to 0$. The zero temperature limit leads to frozen $\sigma = s_0(y)$ at the value $s_0$ minimizing ($\beta > 0$) or maximizing ($\beta < 0$) the exponent $\frac{\pi s^2}{2y} - \mu(s)$ over the support of $\mu$. More generally, positive temperatures encourage larger $\sigma$ values to gather at smaller radii (larger $1/2y$), while negative temperatures encourage them to gather at larger radii (smaller $1/2y$) \cite{TDB2014}.

The constraint equation (\ref{B5}) for the conserved integral $g(s)$ takes the explicit form
\begin{equation}
g(s) = -\frac{\delta F_\sigma}{\delta \mu(s)}
= \int_D d{\bm \rho} p_\sigma(s,y)
\label{5.10}
\end{equation}
Using independence of the two fields in the decoupled model, one obtains trivially
\begin{eqnarray}
\tilde g(s) &=& \int_D d{\bm \rho} \langle \xi({\bm \rho})
\delta[s - \sigma({\bm \rho})] \rangle
\nonumber \\
&=& \int_D d{\bm \rho} \langle \xi({\bm \rho}) \rangle
p_\sigma(s,y) 
\nonumber \\
&\equiv& 0
\label{5.11}
\end{eqnarray}
since $E_G[\xi]$ is a positive even functional of $\xi$.

\begin{figure*}
\includegraphics[height=2.4in,viewport=20 0 440 280,clip]{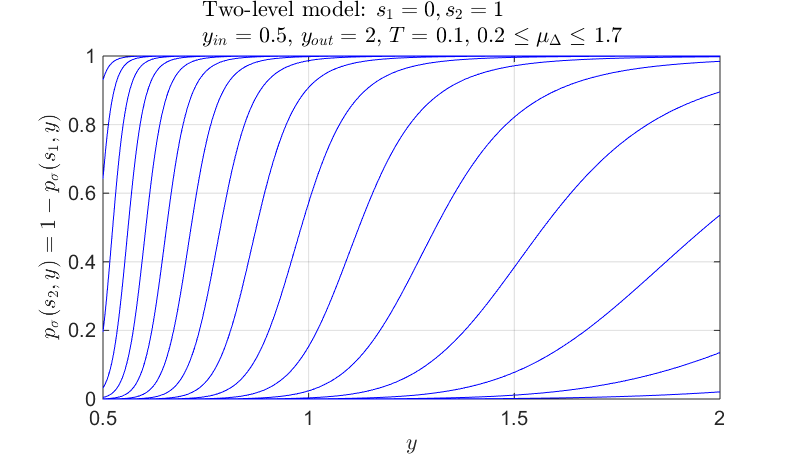}
\quad
\includegraphics[height=2.28in,viewport=10 0 390 280,clip]{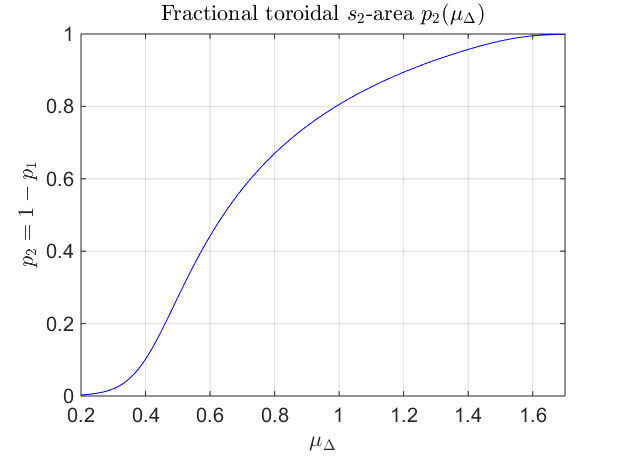}

\caption{Example equilibrium results for the two-level model described by equations (\ref{5.12})--(\ref{5.17}) using $s_1 = 0$ and $s_2 = 1$, scaled temperature $\bar T = 0.1$, and cylinder boundaries $r_\mathrm{in} = 1$, $r_\mathrm{out} = 2$ yielding $y_\mathrm{in} = 0.5$, $y_\mathrm{out} = 2$. \textbf{Left:} Probability distribution $p_\sigma(s_2,y)$ for a range of chemical potential values $0.2 \leq \mu_\Delta \leq 1.7$ in steps of 0.1. The interface moves left (larger region occupied by $s_2$) for increasing $\mu_\Delta$. For the chosen $s_1,s_2$ values, the curves coincide with the mean toroidal flow profile $\langle \sigma({\bm \rho}) \rangle$ [third equality in (\ref{5.14})]. \textbf{Right:} Fractional area $p_2(\mu_\Delta)$, defined by (\ref{5.12}) and (\ref{5.17}), occupied by $\sigma = s_2$. This illustrates the 1--1 correspondence between the Lagrange multiplier $\mu_\Delta$ and the conserved integrals $g(s)$.}

\label{fig:2level_sigma}
\end{figure*}

\subsection{Finite-level models}
\label{sec:finitelevel}

A common approximation is to restrict the initial condition for $\sigma$ to a finite, discrete set of levels. As the simplest model, which will form the basis for most explicit examples in later sections, the two level system is described by
\begin{equation}
g(s) =  A_D [p_1 \delta(s-s_1) + p_2 \delta(s-s_2)]
\label{5.12}
\end{equation}
with, by convention, $s_2 > s_1$, and in which $A_D = \int_D d{\bm \rho} = V_{\cal D}/2\pi$ is the modified coordinate area of $D$, and $p_1, p_2 = 1-p_1$ are, respectively, the fractional areas on which $\sigma({\bm \rho}) = s_1,s_2$. The corresponding chemical potential must take the general form
\begin{equation}
e^{\bar \beta \mu(s)} =  e^{\bar \beta \mu_1} \delta(s-s_1)
+ e^{\bar \beta \mu_2} \delta(s-s_2),
\label{5.13}
\end{equation}
which leads to
\begin{eqnarray}
Z_1 &=& e^{\bar \beta (\mu_1 - \pi s_1^2/2y)}
+ e^{\bar \beta (\mu_2 - \pi s_2^2/2y)}
\nonumber \\
p_\sigma(s_1,y) &=& 1 - p_\sigma(s_2,y)
\nonumber \\
&=& \frac{e^{\bar\beta[\pi(s_2^2 - s_1^2)/4y - \mu_\Delta]}}
{2\cosh\{\bar\beta[\pi(s_2^2 - s_1^2)/4y - \mu_\Delta]\}}
\nonumber \\
\langle \sigma({\bm \rho}) \rangle
&=& s_1 p_\sigma(s_1,y) + s_2 p_\sigma(s_2,y)
\nonumber \\
F_\sigma &=& \frac{\pi (s_1^2 + s_2^2)}{4} I_D - \mu_0 A_D
\label{5.14} \\
&-& \frac{1}{\bar \beta} \int_D d{\bm \rho} \ln\left(2\cosh
\left\{\bar\beta \left[\frac{\pi(s_2^2 - s_1^2)}{4y}
- \mu_\Delta \right] \right\} \right)
\nonumber
\end{eqnarray}
in which we define the mean and difference chemical potentials
\begin{equation}
\mu_0 = \frac{\mu_2 + \mu_1}{2},\ \ \mu_\Delta = \frac{\mu_2 - \mu_1}{2}
\label{5.15}
\end{equation}
and the integration constant
\begin{equation}
I_D = \int_D \frac{d{\bm \rho}}{y}.
\label{5.16}
\end{equation}

The constraint equation (\ref{5.10}) takes the form
\begin{equation}
p_1 = 1 - p_2 = \frac{1}{2A_D} \int d{\bm \rho} p_\sigma(s_1,y).
\label{5.17}
\end{equation}
These depend only on the difference $\mu_\Delta$, which is then used to set the values of $p_{1,2}$. The correspondence is easily seen to be one-to-one and invertible. Some model results are plotted in Fig.\ \ref{fig:2level_sigma}. For example, at low temperatures it is seen that there is sharp (Fermi surface-like) interface between the smaller of $|s_1|,|s_2|$ ($y < y_I$) and the larger of $|s_1|,|s_2|$ ($y > y_I$) dominated regions, with position $y_I = \pi(s_2^2 - s_1^2)/4\mu_\Delta$ controlled by $\mu_\Delta$.

The more general $L$-level model, which restricts $\sigma({\bm \rho})$ to a discrete set of values $s_1 < s_2 < \ldots < s_L$, is described by
\begin{eqnarray}
g(s) &=&  A_D \sum_{m=1}^L p_m \delta(s-s_m),\ \sum_{m=1}^L p_m = 1
\nonumber \\
e^{\bar \beta \mu(s)} &=& \sum_{m=1}^L e^{\bar \beta \mu_m} \delta(s-s_m)
\label{5.18}
\end{eqnarray}
which leads to
\begin{eqnarray}
Z_1 &=& \sum_{m=1}^L e^{-\bar \beta \pi s_m^2/2y}
e^{\bar \beta \mu_m}
\nonumber \\
F_\sigma &=& - \frac{1}{\bar \beta} \int_D d{\bm \rho}
\ln\left(\sum_{m=1}^L e^{\bar \beta \mu_m}
e^{-\bar \beta \pi s_m^2/2y}\right).
\label{5.19}
\end{eqnarray}
Defining $m_0 = \arg \min\{|s_m|\}$ (the index of the smallest magnitude $|s_m|$), the constraint equations may be put in the form
\begin{eqnarray}
p_m &=& \frac{1}{A_D} \int_D d{\bm \rho} p_\sigma(s_m,y)
\label{5.20} \\
p_\sigma(s_m,y) &=& \frac{e^{\bar \beta \Delta \mu_m}
e^{-\bar \beta \pi (s_m^2 - s_{m_0}^2)/2y}}
{1 + \sum_{m' \neq m_0}^L e^{\bar \beta \Delta \mu_{m'}}
e^{-\bar \beta \pi (s_{m'}^2 - s_{m_0}^2)/2y}},
\nonumber
\end{eqnarray}
with the differences $\Delta \mu_m = \mu_m - \mu_{m_0}$, $m \neq m_0$, used to set the $L-1$ independent $p_m$ values. The areas $A_m = A_D p_m$ on which $\sigma = s_m$ are conserved by the flow, though the corresponding domain geometries strongly mix at ever finer scales as time advances.

Analogous to the result for the two-level model, each difference $\Delta \mu_m$ controls the position of an interface, at position $y_m = (s_m^2 - s_{m_0}^2)/2\Delta \mu_m$, separating $s_m$-present ($y > y_m$) and $s_m$-absent ($y < y_m$) regions.

\subsection{Statistics of $\xi$}
\label{sec:statxi}

Since $E_G[\xi]$ is positive definite, (\ref{4.1}) makes sense only for positive temperatures, $\beta > 0$. This is in contrast to the Euler equation \cite{M1990,MWC1992}, where the equivalent of the field $\xi$ is directly constrained by the conservation laws, and negative temperature equilibria are ubiquitous.

In order to verify that the combination $\bar \beta$ is appropriate here as well, consider the square grid Gaussian form
\begin{equation}
e^{-\beta E_G[\xi]} = e^{-\frac{1}{2} \sum_{l,m} A_{lm} \xi_l \xi_m}
\label{5.21}
\end{equation}
with discrete variables
\begin{eqnarray}
\xi_l &=& \xi({\bm \rho}_l)
\nonumber \\
A_{lm} &=& \bar \beta a^2 G({\bm \rho}_l,{\bm \rho}_m).
\label{5.22}
\end{eqnarray}
The discrete approximation to the Green function equation (\ref{2.14}) takes the form
\begin{equation}
- \sum_n \Delta_{ln} G({\bm \rho}_n,{\bm \rho}_m) = \delta_{lm}
\label{5.23}
\end{equation}
in which
\begin{eqnarray}
\Delta_{lm} &=& a^2 [\Delta_*]_{lm}
\nonumber \\
&=& \delta_{l+\hat {\bf y},m} - 2\delta_{lm} + \delta_{l-\hat {\bf y},m}
\nonumber \\
&&+\ \frac{\delta_{l+\hat {\bf z},m} - 2\delta_{lm} + \delta_{l-\hat {\bf z},m}}{2y_l}
\label{5.24}
\end{eqnarray}
is the discrete (nearest neighbor finite difference) version of the generalized Laplacian. The Dirichlet boundary conditions on $G$ lead to the vanishing of (\ref{5.23}) when either or both $l,m$ lie on the boundary, hence (\ref{5.23}) only makes sense if both indices $l,m$ correspond to interior points of $D$. Thus, all terms in (\ref{5.24}) vanish if $l$ itself is a boundary point; and if $l$ is an interior point, the corresponding nearest neighbor Kronecker delta is absent if $l + \hat {\bm \delta}$ is a boundary point (with $\hat {\bm \delta}$ any of the four nearest neighbor directions).

Put another way, from the fact that boundary values of $\xi$ do not contribute to $E_G$ it follows that $\sum_m G_{lm} \xi_m \equiv 0$ for any $\xi$ entirely supported on the boundary. Thus, the matrix $G$ is not invertible unless restricted to the interior. It follows also that, in order to obtain a finite well defined free energy, the functional integral over $\xi$ must be restricted to interior points as well. Note that since the $\sigma$ field is strongly regularized, whether or not one includes values on the boundary is of negligible consequence in the continuum limit, but for consistency we will restrict it to interior sites as well.

\subsubsection{Poloidal field correlations}
\label{subsec:poloidalcorr}

With this interior restriction, $-\Delta_{lm}$ is precisely the inverse of the matrix $G_{lm}$, and the second moments of $\xi$ are given by the corresponding nearest neighbor form
\begin{equation}
\langle \xi_l \xi_m \rangle = [A^{-1}]_{lm}
= -\frac{\Delta_{lm}}{\bar \beta a^2}
= -\frac{1}{\bar \beta} [\Delta_*]_{lm}.
\label{5.25}
\end{equation}
The right hand side diverges in the continuum limit: microscopic vorticity fluctuations are extremely large. Note that, when expressed in terms of $\beta = \bar \beta/a^2$, (\ref{5.25}) may be expressed in the form
\begin{equation}
\langle \xi({\bm \rho}) \xi({\bm \rho}') \rangle
= -\frac{1}{\beta} \Delta_* \delta({\bm \rho}-{\bm \rho}'),
\label{5.26}
\end{equation}
which is well defined in the sense of distributions. However, for finite $\beta$ this leads to much stronger divergence of the variance $\langle \xi({\bm \rho})^2 \rangle \propto 1/\beta a^4$ that will be seen to lead to unphysical states with infinite flow energy.

By substitution of (\ref{2.18}), it is straightforward to verify the stream function correlation identity
\begin{equation}
\langle \psi_D({\bm \rho}) \psi_D({\bm \rho}') \rangle
= \frac{1}{\beta} G({\bm \rho},{\bm \rho}').
\label{5.27}
\end{equation}
The result is finite for finite $\beta$, but the remaining short-range singularity in $G$ still leads to a logarithmic divergence $\langle \psi_D({\bm \rho})^2 \rangle \sim \ln(A_D/a^2)$. Similarly, the microscale mean square difference $\langle [\psi_D({\bm \rho}) - \psi_D({\bm \rho}')]^2 \rangle \sim \beta^{-1} \ln(|{\bm \rho} - {\bm \rho}'|/a)$ increases logarithmically with separation. Both of these results reflect the thermally rough elastic surface model that emerges when $E_G$ is written in terms of the stream function [gradient-squared form in the second line of (\ref{3.1})].

On the other hand, it is straightforward to verify that the mean energy $\langle E_G \rangle \sim A_D/2 \beta a^2$ also diverges for finite $\beta$. A consistent, finite conserved value of the hydrodynamic kinetic energy emerges only if the combination $\bar \beta = \beta a^2$ remains finite, consistent with the scaling (\ref{5.8}) that emerged from the statistics of $\sigma$. With this scaling, one sees that $\langle \psi_D({\bm \rho}) \psi_D({\bm \rho}') \rangle \to 0$: the macroscopic stream function vanishes. On the other hand, consistent with finite kinetic energy, the continuum limit correlations of the 2D velocity ${\bf v}_D = (v_r,v_z) = (-\partial_z \psi_D/\sqrt{2y}, \partial_y \psi_D)$ take the form
\begin{equation}
\langle {\bf v}_D({\bm \rho}) \cdot {\bf v}_D({\bm \rho}') \rangle
= \frac{1}{\beta} \delta({\bm \rho}-{\bm \rho}')
= \frac{1}{\bar \beta} \delta_{{\bm \rho},{\bm \rho}'}
\label{5.28}
\end{equation}
and are therefore finite. Although $\psi_D$ vanishes, and any local average of ${\bf v}_D$ vanishes as well, the finite microscale gradients of the former generate finite microscale velocity fluctuations containing finite energy. This is in contrast to the Euler case in which $\xi$ is directly bounded, ${\bf v}_D$ is then continuous, $\psi$ is differentiable, and the energy resides entirely in the macroscale flow. There are parallels in the present case to that of magnetohydrodynamic equilibria \cite{W2012}, where the infinite number of integral constraints apply to the flow potentials rather than to the vorticity: finite energy again provides the essential constraint on vorticity fluctuations, and microscale velocity fluctuations make a finite contribution to the energy.

\subsubsection{Poloidal field free energy}
\label{subsec:poloidalF}

In a similar fashion, the poloidal free energy contribution is
\begin{eqnarray}
F_\xi(\bar \beta) &=& \frac{1}{2\beta} \ln [\det({\bf A}/2\pi)]
\nonumber \\
&=& -\frac{1}{2\beta} \sum_n \ln\left(\frac{2\pi a^2 \lambda_n}{\bar \beta} \right)
\label{5.29}
\end{eqnarray}
in which $\lambda_n$ are the eigenvalues of the generalized Poisson equation
\begin{equation}
-\sum_m \Delta_{lm} \psi_m = \lambda a^2 \psi_l \ \to \
-\Delta_* \psi = \lambda \psi.
\label{5.30}
\end{equation}
with the usual Dirichlet and periodic boundary conditions.

For the cylindrical geometry, translation invariance in $z$ allows one to seek eigenfunctions of (\ref{5.30}) in the form
\begin{equation}
\psi({\bm \rho}) = e^{iqz} \psi_q(y),\ \ q = \frac{2\pi m}{h}
\label{5.31}
\end{equation}
for integer $|m| \leq h/2a$. For each $q$ one then solves the 1D eigenvalue equation \cite{AS1970}
\begin{equation}
\left(-\partial_y^2 + \frac{q^2}{2y} \right) \psi_q = \lambda \psi_q
\label{5.32}
\end{equation}
with Dirichlet boundary conditions on the interval $[y_\mathrm{in}, y_\mathrm{out}]$. The eigenvalues $\lambda_{q,l}$ are substituted into (\ref{5.29}) with $n = (q,l)$ now a double index.

For finite $a$ there are $N_D = A_D/a^2$ eigenvalues, and one expects a finite limit
\begin{equation}
\Lambda_0 = \lim_{a \to 0} a^2 \sum_n \ln(a^2 \lambda_n),
\label{5.33}
\end{equation}
The continuum free energy therefore takes the form
\begin{equation}
F_\xi(\bar \beta) = \frac{1}{2\bar \beta} [\ln(\bar \beta/2\pi ) - \Lambda_0],
\label{5.34}
\end{equation}
and is finite for finite $\bar \beta$. Note that the Dirichlet boundary conditions on the eigenfunctions, which ensures that all $\lambda_n > 0$, automatically remove the boundary values of $\xi$ from the functional integral, rendering the free energy well defined [as per the discussion below (\ref{5.24})].

To gain intuition using an analytic example, if one replaces $2y_l$ by unity in (\ref{5.24}), (\ref{5.30}) and (\ref{5.32}) \cite{foot:gridunits}, the Fourier representation is appropriate in both $y$ and $z$, and one obtains eigenfunctions and eigenvalues
\begin{eqnarray}
\psi({\bm \rho}) &=& \psi_{pq} e^{i(p y + q z)},\ \
(p,q) = 2\pi \left(\frac{l}{y_\mathrm{out} - y_\mathrm{in}},\frac{m}{h} \right)
\nonumber \\
\lambda(p,q) &=& \frac{\sin^2(p a/2) + \sin^2(q a/2)}{a^2/4},
\label{5.35}
\end{eqnarray}
for integers $|l| \leq (y_\mathrm{out} - y_\mathrm{in})/2a$, $|m| \leq h/2a$. This leads, as stated, to the finite limit
\begin{eqnarray}
\Lambda_0 &=& A_D \int_{-\pi}^\pi \frac{ds_y}{2\pi}
\int_{-\pi}^\pi \frac{ds_z}{2\pi}
\nonumber \\
&&\times\ \ln\left[4\sin^2(s_y/2) + 4\sin^2(s_z/2) \right].
\label{5.36}
\end{eqnarray}
Note that the scaling (\ref{5.30}) yields the finite result $\lambda({\bf q}) \to |{\bf q}|^2$ as $a \to 0$ for any finite ${\bf q}$---and will similarly yield a well defined functional form when the $2y$ factor is restored in the eigenvalue equation. However, the free energy includes contributions from all scales, and the full microscale form of the eigenvalues enters $\Lambda_0$. As a consequence, the value of $\Lambda_0$ depends on the precise form of the discretization. On the other hand, the $\ln(\bar \beta/2\pi)$ term is universal, and contains the essential temperature-dependent thermodynamic behavior, which is independent of the precise form of the continuum limit.

\subsubsection{Stream function representation}
\label{subsec:streamfn}

Note also that one may use the eigen-decomposition to perform the phase space change of variable
\begin{equation}
\int D[\xi] = J_\xi \int D[\psi]
\label{5.37}
\end{equation}
with (constant) Jacobian
\begin{equation}
J_\xi = \det\left[\frac{\partial \xi}{\partial \psi} \right]
= \prod_n \lambda_n.
\label{5.38}
\end{equation}
An alternative form for the partition function is therefore
\begin{eqnarray}
Z_\xi &=& J_\xi \int D[\psi] e^{-\beta E_G[\psi]}
\nonumber \\
E_G[\psi] &=& -\frac{1}{2} \int d{\bm \rho}\, \psi \Delta_* \psi
= \frac{1}{2} \int d{\bm \rho} |\nabla_* \psi|^2.\ \ \ \ \ \
\label{5.39}
\end{eqnarray}
Using the eigen-decomposition to diagonalize $E_G[\psi]$, the free energy is
\begin{equation}
F_\xi(\bar \beta) = -\frac{1}{\beta} \ln(J_\xi)
+ \frac{1}{2\beta} \sum_n \ln(\beta \lambda_n/2\pi),
\label{5.40}
\end{equation}
which, using (\ref{5.33}), reproduces (\ref{5.29}) and (\ref{5.34}).

\subsection{Decoupled model thermodynamics}
\label{sec:TDdecoupled}

We have noted that $F_\sigma$ is well defined for both positive and negative temperatures [assuming only bounded support of $g(s)$], while $F_\xi$ [see (\ref{5.34})] is well defined only for positive temperatures. Physically, the toroidal flow energy
\begin{equation}
E_\sigma(\bar \beta, \mu) = \int_D d{\bm \rho}
\int ds p_\sigma(s,y) \frac{\pi s^2}{2y}
\label{5.41}
\end{equation}
is always finite, while the poloidal flow energy
\begin{equation}
E_\xi(\bar \beta) = \frac{\partial (\bar \beta F_\xi)}{\partial \bar \beta}
= \frac{1}{2} \bar T
\label{5.42}
\end{equation}
(a version of the equipartition principle for quadratic Hamiltonians) diverges as $\bar \beta \to 0^+$ ($\bar T \to +\infty$), and is strictly infinite for $\bar \beta < 0$. It follows that arbitrarily large values of the total energy
\begin{equation}
E_0 = E_\sigma(\bar \beta, \mu) + \frac{1}{2} \bar T
\label{5.43}
\end{equation}
are explored using only positive values of $\bar T$, with the toroidal energy saturating at the fully mixed value
\begin{equation}
E_\sigma(0, \mu) = \frac{\pi I_D}{2 A_D} \int s^2 ds g(s),
\label{5.44}
\end{equation}
in which one observes that $\lim_{\bar \beta \to 0} p_\sigma(s,y) = g(s)/A_D$ independent of $y$. All remaining energy is absorbed into ever increasing $\xi$ fluctuations. Negative temperature toroidal states are never accessed by the combined system.

\section{General coupled model}
\label{sec:genmodel}

We consider now the full coupled model (\ref{4.1}). The $\xi \tilde \mu(\sigma)$ coupling term is sufficiently simple that it is straightforward to perform the functional integral over either one of the fields to obtain a reduced model expressed entirely in terms of the other. The resulting models have more complex interactions, and an exact solution is not generally possible. However, general features may be understood, and approximate solutions may be derived in various limits.  In fact, it will turn out that the proper $a \to 0$ scaling limit for the coupling requires that $\tilde \mu = \bar \mu a^2$ vanish with $a$, allowing an exact relation between $\tilde \mu$ and $\tilde g$ to be derived.

Both reduced models must produce the identical final free energy (\ref{4.2}), hence represent the same underlying physics, though expressed in different ways. The effective (conditional) equilibrium statistics of the fields $\sigma$ and $\xi$ provide interesting complementary views of the underlying fluctuations:

\begin{enumerate}
\item The $\sigma$-model is a classical scalar spin model with local (nearest neighbor) antiferromagnetic interactions, but including also adjustable on-site potentials determined by the Lagrange multipliers $\mu,\tilde \mu$ that allow one to enforce the conservation laws (\ref{B5}) [with (\ref{3.15})].

\item The $\xi$-model continues to be a vortex model with long-range Coulomb-type interactions, but now with an additional local potential. The potential is (linearly) unbounded from below, hence does not confine the $\xi$ field to finite values as would be required for validity of the variational approach---the variance of $\xi$ is still $O(1/\bar \beta a^2)$. The forms of $\mu,\tilde \mu$ that determine the exact shape of the potential again allow one to enforce the conservation laws.
\end{enumerate}

\subsection{Reduced $\sigma$-model}
\label{sec:sigma_red}

We consider first the case in which the (Gaussian) $\xi$ integral is performed to obtain an effective model for $\sigma$ alone. We apply the Gaussian identity
\begin{eqnarray}
&&\prod_l \int_{-\infty}^\infty d\xi_l e^{\sum_l q_l \xi_l}
e^{-\frac{1}{2} \sum_{l,m} A_{lm} \xi_l \xi_m}
\nonumber \\
&&\ \ \ \ \ =\ \frac{1}{\mathrm{\sqrt{det(2\pi {\bf A})}}}
e^{\frac{1}{2} \sum_{l,m} [{\bf A}^{-1}]_{lm} q_l q_m},
\label{6.1}
\end{eqnarray}
valid for any positive definite (real symmetric) matrix ${\bf A}$. In addition to the parameters (\ref{5.22}), we identify here
\begin{equation}
q_l = \bar \beta \{\tilde \mu[\sigma({\bm \rho}_l)] + h_\xi({\bm \rho}_l) \},
\label{6.2}
\end{equation}
in which ${\bm \rho}_l$ is again restricted to the interior of $D$, and the conjugate field $h_\xi$ has also been reinstated.

Restoring continuum notation, it follows that the result of the $\xi$ functional integral is
\begin{equation}
Z = \frac{e^{-\beta E_z^0}}{\sqrt{\mathrm{det}(2\pi\beta a^4 G)}}
\int D[\sigma] e^{-\beta {\cal K}_1[\sigma]}
\label{6.3}
\end{equation}
with reduced $\sigma$-functional
\begin{eqnarray}
{\cal K}_1[\sigma] &=& \int_D d{\bm \rho}
\left\{\frac{\pi \sigma^2}{2y} - \mu(\sigma) - h_\sigma \sigma \right.
\nonumber \\
&&\ \ \ \ \ +\ \left. \frac{1}{2}
[\tilde \mu(\sigma) + h_\xi] \Delta_*
[\tilde \mu(\sigma) + h_\xi] \right\}. \ \ \ \ \
\label{6.4}
\end{eqnarray}
When interpreting the last term, the interior restriction allows one to take $\tilde \mu(\sigma)$ and $h_\xi$ to vanish on the boundary. Note that the term
\begin{eqnarray}
\tilde M &\equiv& -\frac{1}{2} \beta \int_D d{\bm \rho}
\tilde \mu(\sigma) \Delta_* \tilde \mu(\sigma)
\nonumber \\
&=& \frac{1}{2} \beta \int_D d{\bm \rho}
\tilde \mu'(\sigma)^2 |\nabla_* \sigma|^2
\label{6.5}
\end{eqnarray}
for $\beta > 0$ \emph{favors} gradients in $\sigma$, corresponding to \emph{antiferromagnetic} correlations at the grid scale. This is converse to the \emph{ferromagnetic} negative temperature states encountered for the Euler equation \cite{MWC1992}, which favor smooth $\sigma$.

\subsubsection{Scaling of $\tilde \mu(\sigma)$}
\label{subsec:muscale}

In discrete form one obtains
\begin{eqnarray}
\beta {\cal K}_1[\sigma] &=& \bar \beta \sum_l \left[
\frac{\pi \sigma_l^2}{2y_l} - \mu(\sigma_l) - h_{\sigma,l} \sigma_l \right]
\label{6.6} \\
&&+\ \frac{\bar \beta}{2a^2} \sum_{l,m} \Delta_{lm}
[\tilde \mu(\sigma_l) + h_{\xi,l}] [\tilde \mu(\sigma_m) + h_{\xi,m}]
\nonumber
\end{eqnarray}
The first line generates the individual spin weighting analyzed in Sec.\ \ref{sec:statsig}. The second line generates antiferromagnetic nearest neighbor interactions between spins. Note the appearance of the divergent coefficient $\beta = \bar \beta/a^2$, rather than $\bar \beta$, in the latter. We will see below that in order to obtain the proper scaling that produces finite values for the conserved integrals $\tilde g(s)$, one requires an additional scaling relation
\begin{equation}
\tilde \mu(\sigma) = a \bar \mu_0 + a^\gamma \bar \mu(\sigma),\ \
h_\xi = a^\gamma \bar h_\xi
\label{6.7}
\end{equation}
for some $\gamma > 0$, with $\bar \mu$ and $\bar h_\xi$ remaining finite as $a \to 0$. From (\ref{6.5}) one sees that $\bar \mu_0$ contributes only at the boundary, but is required, with its potentially distinct scaling, to control the total vorticity $\int_D d{\bm \rho} \xi({\bm \rho})$. The choice $\gamma = 1$ produces a finite antiferromagnetic coupling in (\ref{6.6}). We will see that this leads to divergent $\tilde g(s)$, and a larger value $\gamma = 2$, hence asymptotically vanishing coupling, is required.

Physically, the coupling biases the equilibrium state toward differing neighboring toroidal flows $\sigma$, beyond that which would be encountered with pure random (Poisson statistics) assignment, $\tilde \mu \equiv 0$.  This is precisely what is required to control the specified nonzero average $\tilde g(s)$ of $\xi$ over different level sets $\{\sigma = s\}$ [see (\ref{3.15}) and (\ref{B5})]. Since real fluids do not have a fixed microscale $a$, it is not clear how such correlations would actually be exhibited in an equilibrating fluid, especially as finite viscosity will lead to at some point to significant dissipation. As discussed in more detail in Sec.\ \ref{sec:conclude}, this would be an interesting topic for future numerical investigation.

\subsubsection{Equilibrium averages}
\label{subsec:eqave}

The equilibrium average of $\xi$ follows in the form
\begin{eqnarray}
\xi_\mathrm{eq}({\bm \rho}) &\equiv& \langle \xi({\bm \rho}) \rangle
= -\frac{\delta F_0}{\delta h_\xi({\bm \rho})}
= -\langle \Delta_* \tilde \mu[\sigma({\bm \rho})] \rangle
\nonumber \\
&=& -\Delta_* \psi_\mathrm{eq}({\bm \rho})
\nonumber \\
\psi_\mathrm{eq}({\bm \rho}) &\equiv&
\langle \tilde \mu[\sigma({\bm \rho})] \rangle,
\label{6.8}
\end{eqnarray}
in which the averages are now with respect to the reduced functional ${\cal K}_1[\sigma]$, and we set $h_\xi = 0$ at the end. The adopted convention $\tilde \mu[\sigma({\bm \rho})] \equiv 0$ on $\partial D$ ensures that $\psi_\mathrm{eq}$ obeys the required Dirichlet boundary conditions.  Although $\tilde \mu(\sigma)$, like $\sigma$ itself, will have strong microscale fluctuations, $\psi_\mathrm{eq}({\bm \rho})$ will be a smooth function in the interior of $D$, and it follows that $\langle \xi({\bm \rho}) \rangle$ will be as well. The scaling (\ref{6.7}) actually results in $\psi_\mathrm{eq} \to 0$, while
\begin{equation}
\xi_\mathrm{eq}({\bm \rho}) \to \bar \mu_0
[\delta(y-y_\mathrm{in}) + \delta(y-y_\mathrm{out})],
\label{6.9}
\end{equation}
corresponding to vanishing interior vorticity, but a pair of uniform finite vortex surface layers with total mean vorticity
\begin{equation}
\xi_0 = \frac{1}{A_D} \int_D d{\bm \rho} \xi_\mathrm{eq}({\bm \rho})
= \frac{2 \bar \mu_0}{y_\mathrm{out} - y_\mathrm{in}}.
\label{6.10}
\end{equation}
Unequal vortex layers generate a net uniform flow, and hence provide an equivalent mechanism for producing the conserved vertical flow $v_z^0$. Equal layers produce zero net interior flow, hence vanishing energy contribution. Physically, this is directly analogous to the usual Coulomb result that all excess charge on a conducting body resides on the surface in such a way that the interior electric field vanishes. Note that for $\gamma > 1$ the stream function fluctuations $\psi_D \sim a/\sqrt{\bar \beta}$ are much larger than those in $\tilde \mu(\sigma)$, and produce the finite microscale velocity fluctuations \cite{foot:AFpsi}.

By way of contrast, if spread uniformly, $\xi({\bm \rho}) \equiv \xi_0$, the result is a linear shear flow
\begin{eqnarray}
\psi_\mathrm{eq}(y) &=& \frac{1}{2} \xi_0 (y-y_\mathrm{in})(y_\mathrm{out}-y)
\nonumber \\
v_\mathrm{eq}^z(y) &=& -\xi_0
\left(y - \frac{y_\mathrm{in} + y_\mathrm{out}}{2} \right).
\label{6.11}
\end{eqnarray}
However, in the context of the Euler equation, this corresponds to a ``zonal jet'' negative temperature state, and is in the present case thermodynamically unstable to the positive temperature state (\ref{6.10}).

The conserved integrals (\ref{B5}) follow in the form
\begin{eqnarray}
g(s) &=& -\frac{\delta F_0}{\delta \mu(s)}
= \int_D d{\bm \rho} \langle \delta[s - \sigma({\bm \rho})] \rangle
\label{6.12} \\
\tilde g(s) &=& -\frac{\delta F_0}{\delta \tilde \mu(s)}
= - \int_D d{\bm \rho} \langle \Delta_* \tilde \mu[\sigma({\bm \rho})]
\delta[s - \sigma({\bm \rho})] \rangle
\nonumber
\end{eqnarray}
in which we have set $h_\xi = h_\sigma = 0$. The substitution $\xi \to -\Delta_* \tilde \mu(\sigma)$ is consistent with (\ref{6.8}). However, although the mean (\ref{6.8}) vanishes,  the confinement of the integration support to a particular level set $\sigma = s$ in general biases the integrand defining $\tilde g(s)$ to produce a nonzero result. For example, if $s$ is at the high end of the support of $g(s)$, then the neighboring points ${\bm \rho} \pm a \hat {\bf y}$, ${\bm \rho} \pm a \hat {\bf z}$ will likely lie on lower level sets, biasing the finite difference Laplacian $a^2 \Delta_* \sigma({\bm \rho})$ [see (\ref{5.24})] to finite negative values, and $a^2 \Delta_* \bar \mu[\sigma({\bm \rho})]$ to (also typically finite) values depending on the precise form of the function $\bar \mu(s)$.

It follows that finite biased values of $\Delta_* \tilde \mu[\sigma({\bm \rho})] = a^{\gamma - 2} (a^2 \Delta_*) \bar \mu[\sigma({\bm \rho})]$ require the choice
\begin{equation}
\gamma = 2,
\label{6.13}
\end{equation}
as claimed above. This choice confirms, via (\ref{6.8}), that the bias in $\xi$ when ${\bm \rho}$ is confined to a particular level set of $\sigma$ is also finite, as required by the original form (\ref{B5}) with (\ref{3.15}).

The total vorticity is
\begin{equation}
A_D \xi_0 = \int ds \tilde g(s) = 2 h \bar \mu_0,
\label{6.14}
\end{equation}
and is nonzero only by virtue of the surface layers.

\subsubsection{Exact relations for $g(s)$ and $\tilde g(s)$}
\label{subsec:gexact}

With the choice (\ref{6.13}), the last line of the statistical functional (\ref{6.6}) is of relative order $a^2$ compared to the first, and it follows that the averages (\ref{6.12}) may be evaluated using the first term alone, i.e., the uncoupled model---the additional antiferromagnetic bias (beyond the appearance of $\tilde \mu$ in the integrand of that equation) is negligible compared to that induced by the choice of $s$. This produces the following exact solution.

The conserved integral $g(s)$ is still given by (\ref{5.9}), with inputs (\ref{5.6}) and (\ref{5.5}).

Moving on to $\tilde g(s)$, since all sites are completely independent in the decoupled model, for ${\bm \rho} \neq {\bm \rho}'$, even for nearest neighbor microscale grid sites, one obtains
\begin{eqnarray}
\langle \bar \mu[\sigma({\bm \rho}')]
\delta[\sigma({\bm \rho}) - s] \rangle_0
&=& \langle \bar \mu[\sigma({\bm \rho}')] \rangle_0
\langle \delta[\sigma({\bm \rho}) - s] \rangle_0
\nonumber \\
&=& p_\sigma(s,y) \int ds' \bar \mu(s') p_\sigma(s',y'),
\nonumber \\
\label{6.15}
\end{eqnarray}
while, for ${\bm \rho} = {\bm \rho}'$,
\begin{eqnarray}
\langle \bar \mu[\sigma({\bm \rho})]
\delta[\sigma({\bm \rho}) - s] \rangle_0
&=& \bar \mu(s) \langle \delta[\sigma({\bm \rho}) - s] \rangle_0
\nonumber \\
&=& \bar \mu(s) p_\sigma(s,y)
\label{6.16}
\end{eqnarray}
in which $\langle \cdot \rangle_0$ denotes the decoupled model average, $\tilde \mu \equiv 0$, and the single site probability $p_\sigma$ is defined by (\ref{5.6}). Substituting the discrete form (\ref{5.22}), along with the scaling (\ref{6.7}) and (\ref{6.13}), one obtains
\begin{eqnarray}
\Delta_* \tilde \mu(\sigma_l)
&\to& \bar \mu(\sigma_{l + \hat {\bf y}})
+ \bar \mu(\sigma_{l - \hat {\bf y}})
+ \frac{\bar \mu(\sigma_{l + \hat {\bf z}})
+ \bar \mu(\sigma_{l - \hat {\bf z}})}{2y_l}
\nonumber \\
&&-\ \left(2 + \frac{1}{y_l} \right) \bar \mu(\sigma_l),
\label{6.17}
\end{eqnarray}
as long as $l$ does not neighbor a boundary point (i.e, $l \pm \hat {\bf y}$ are both not boundary points), while
\begin{equation}
-\Delta_* \tilde \mu(\sigma_l) \to \frac{\bar \mu_0}{a}
\label{6.18}
\end{equation}
for $l$ neighboring a boundary point.

Substituting (\ref{6.16})--(\ref{6.18}) into the second line of (\ref{6.12}) and restoring continuum limit notation, one obtains
\begin{eqnarray}
\tilde g(s) &=& \int_D d{\bm \rho}  p_\sigma(s,y)
\left(2 + \frac{1}{y} \right)
\nonumber \\
&&\times\ \left[\int ds' \bar \mu(s') p_\sigma(s',y) - \bar \mu(s) \right]
\nonumber \\
&&+\ h\bar \mu_0 [p_\sigma(s,y_\mathrm{in}) + p_\sigma(s,y_\mathrm{out})]
\label{6.19}
\end{eqnarray}
in which, in the first term, one may safely replace $p_\sigma(s,y + a \hat {\bm \delta}) \to p_\sigma(s,y)$ for all $\hat {\bm \delta}$ at the end because ratios of small differences are no longer involved. The $s$-integral of the first term vanishes, while the last line produces (\ref{6.14}).

Equation (\ref{6.19}) [replacing (\ref{5.11})], along with (\ref{5.10}), are the fundamental results of this section. Note that the relation between $\mu$ and $g$ is independent of $\bar \mu$, but strongly nonlinear. However, once $\mu$ is determined, the relation between $\tilde g$ and $\bar \mu$ is linear. Some illustrative applications will be presented in Sec.\ \ref{sec:apps}.

\subsection{Reduced $\xi$-model}
\label{sec:xi_red}

Although the $\sigma$-field formulation (\ref{6.4}) provides the most convenient representation, being directly expressed in terms of the conserved field, it is interesting to examine also the alternative reduced model obtained by integrating out $\sigma$. Since the latter appears without any coupling between different spatial points, we can express the result in terms of the 1D integral
\begin{equation}
e^{\bar \beta W(x,t;h_\sigma,h_\xi)} = \int ds
e^{\bar \beta \{\mu(s) + x [\tilde \mu(s) + h_\xi]
- \pi s^2/2t + h_\sigma s\}},
\label{6.20}
\end{equation}
which yields
\begin{equation}
Z = e^{-\beta E_z^0} \int D[\xi] e^{-\beta {\cal K}_2[\xi]}
\label{6.21}
\end{equation}
with reduced $\xi$-functional
\begin{equation}
{\cal K}_2[\xi] = E_G[\xi]
- \int_D d{\bm \rho} W[\xi({\bm \rho}),y;h_\sigma({\bm \rho}),h_\xi({\bm \rho})].
\label{6.22}
\end{equation}
The positions of the level curves of $\sigma$ fluctuate with the statistical mechanical average, hence the statistics of $\xi$ at each physical point ${\bm \rho}$ involve also an average over the possible $\sigma$-level curves passing through that point. This is precisely the content of the function $W$. As will be discussed below, this model must contain the identical antiferromagnetic interpretation as the $\sigma$ formulation.

The result here is formally similar to that for the Euler equation \cite{MWC1992}, except for that case the range of $\xi$ was directly bounded by the vorticity constraints. Here it is unbounded since the vorticity constraints instead apply to $\sigma$. For example, the two level system form yields
\begin{equation}
W(x,t) = \frac{1}{\bar \beta}
\ln\left[e^{\bar \beta (\mu_1 + x \tilde \mu_1 - \pi s_1^2/2t)}
+ e^{\bar \beta (\mu_2 + x \tilde \mu_2) - \pi s_2^2/2t} \right],
\label{6.23}
\end{equation}
in which $\tilde \mu_l = \tilde \mu(s_l)$, $l=1,2$. This result contains linearly increasing terms for $\mathrm{sgn}(x)$ such that $x \tilde \mu_l > 0$ [as should be more generally clear from the $\xi \tilde \mu(\sigma)$ dependence in (\ref{4.1})]. These are controlled by the positive definite quadratic form $E_G[\xi]$, but as a consequence there remain large fluctuations in equilibrium, and the Euler equation mean field result fails (which is a consequence \emph{both} of the bounded vorticity and of the long-range property of $G$).

In the absence of $W$, as seen in detail in Sec.\ \ref{sec:decoupledmodel}, one has Fourier coefficient $\hat \xi({\bf q}) \sim q/\sqrt{\bar \beta}$, strongly divergent at small scales, $q \sim \pi/a$, leading to $\langle \xi({\bm \rho})^2 \rangle \sim 1/a^4 \to \infty$ [see (\ref{5.26})]. From the definition (\ref{6.20}) one sees that
\begin{equation}
e^{\bar \beta W(x,t)} \sim \left\{
\begin{array}{ll}
e^{\bar \beta \tilde \mu_\mathrm{max} x}, & x \to \infty \\
e^{\bar \beta \tilde \mu_\mathrm{min} x}, & x \to -\infty
\end{array} \right.
\label{6.24}
\end{equation}
in which
\begin{equation}
\tilde \mu_\mathrm{max} = \sup \tilde \mu(s),\ \
\tilde \mu_\mathrm{min} = \inf \tilde \mu(s)
\label{6.25}
\end{equation}
are the maximum and minimum values of $\tilde \mu(s)$. It follows that $W(x,t) \to +\infty$ on at least one side for large $|x|$ (depending on the signs of $\tilde \mu_\pm$), thus enhancing, rather than suppressing, fluctuations of $\xi$. This contrasts with the Euler equation case, in which the analogue of $W \to -\infty$ outside a finite range of support of $\xi$, directly limiting its fluctuations.

As in the decoupled model one therefore relies on the quadratic term $E_G$ to counteract this divergence, and one expects a self-consistent shift in the mean of the form $\delta \xi \sim -\Delta_{\bm \rho} W(\xi,y) \sim (\tilde \mu_\mathrm{max} - \tilde \mu_\mathrm{min})/a^2$, accounting for the site-to-site near-independence. The finite shift requirement leads to the scaling $\tilde \mu_\mathrm{max,min} = a^2 \bar \mu_\mathrm{max,min}$, recovering (\ref{6.7}) and (\ref{6.13}). Note also that the combination $\bar \beta \tilde \mu \xi = \bar \beta \bar \mu \delta \xi a^2 \to 0$ for $\xi \sim 1/a$: the coupling term is a vanishing perturbation of the decoupled model, but is precisely the right size to provide the finite bias to the large $q \sim \pi/a$ Fourier components that dominate the microscale mixing.

The mean value $\tilde g(s)$ of $\xi$ is \emph{fixed} on each $\sigma = s$ level curve, set here through the Lagrange multiplier $\tilde \mu(s)$. However, there is only weak control of fluctuations about this mean, which remain comparable to those for the decoupled model. Thus, the equilibria will be strongly fluctuating and the variational mean field approximation will fail.  The elastic membrane model (\ref{5.39}) defined by the stream function $\psi$ will fluctuate strongly, though remaining microscopically consistent with the mean values defined by $\tilde \mu(s)$ [see (\ref{6.8})] and constrained by the Dirichlet boundary condition on the domain $D$.

From (\ref{B5}), the conserved integrals take the form
\begin{eqnarray}
g(s) &=& \left \langle \int_D d{\bm \rho}
\frac{\delta W[\xi({\bm \rho}),y]}{\delta \mu(s)} \right\rangle
\nonumber \\
&=& \int_D d{\bm \rho} \langle \gamma[s|\xi({\bm \rho})] \rangle
\nonumber \\
\tilde g(s) &=& \left \langle \int_D d{\bm \rho}
\frac{\delta W[\xi({\bm \rho}),y]}{\delta \tilde \mu(s)} \right\rangle
\nonumber \\
&=& \int_D d{\bm \rho} \langle \xi({\bm \rho}) \gamma[s|\xi({\bm \rho})] \rangle
\label{6.26}
\end{eqnarray}
in which the averages are now with respect to the reduced functional ${\cal K}_2[\xi]$, and
\begin{eqnarray}
\gamma[s|\xi({\bm \rho})] &=& \langle \delta[s - \sigma({\bm \rho})] \rangle_\sigma
\nonumber \\
&=& e^{-\bar \beta W[\xi({\bm \rho}),y]}
e^{\bar \beta [\mu(s) + \xi({\bm \rho}) \tilde \mu(s) - \pi s^2/2y]}
\ \ \ \ \ \
\label{6.27}
\end{eqnarray}
is the probability distribution of $s = \sigma({\bm \rho})$ for fixed $\xi({\bm \rho})$ (with $\langle \cdot \rangle_\sigma$ being the average over $\sigma$ at fixed $\xi$). For fixed $s$, (\ref{6.24}) and (\ref{6.25}) control the behavior of $\gamma$ for large $|\xi({\bm \rho})|$.

Using the scaling (\ref{6.7}) and (\ref{6.13}), it is straightforward to rederive the forms (\ref{5.10}) and (\ref{6.19}) for the conserved integrals. Specifically, one may simply drop the $\tilde \mu$ dependence in the expression for $g(s)$---it generates vanishing corrections for $a \to 0$. Similarly, in the expression for $\tilde g$ the $\xi$ dependence in the $e^{-\bar \beta W}$ factor in (\ref{6.27}) may be dropped, and the average of the resulting combination $p_\sigma(s,y) \xi e^{\bar \beta \tilde \mu(s) \xi}$ reproduces the result (\ref{6.19}).

\section{Applications}
\label{sec:apps}

We consider now a few examples, illustrating the results of the theory.

\subsection{Uniform vorticity bias}
\label{sec:uniformvort}

Consider first the case of constant $\tilde \mu(s) = a \bar \mu_0$, in which only the uniform term is kept in (\ref{6.7}). As will be seen, the long range Coulomb interactions make this a rather singular limit. For short range interactions, a finite uniform mean vorticity would be expected [see (\ref{6.11})], but, as we have previously seen, the long range interactions push the extra vorticity to the boundaries [see (\ref{6.9})].

Since dependence of $\sigma$ drops out, this represents a somewhat more general decoupled model \cite{TDB2014}, and it follows that the $\sigma$-field distribution function (\ref{5.6}), free energy (\ref{5.8}), and conserved integrals (\ref{5.9}) are unchanged.

The $\xi$-field partition function, defined by (\ref{6.1}) with constant $q_l = \bar \beta a \bar \mu_0$, leads to the correction
\begin{equation}
F_\xi(\bar \beta, \bar \mu_0) = F_\xi(\bar \beta) -  a h \bar \mu_0^2
\label{7.1}
\end{equation}
in which the first term is given by (\ref{5.34}) and the second term is the result of the area integral of
\begin{equation}
-\frac{1}{2} \tilde \mu \Delta_* \tilde \mu
= \frac{1}{2} a \bar \mu_0^2
[\delta(y - y_\mathrm{in}) + \delta(y - y_\mathrm{out})],
\label{7.2}
\end{equation}
or equivalently the area integral of $\frac{1}{2} |\nabla_* \tilde \mu|^2$. The correction vanishes in the continuum limit, consistent with the physical result that the equilibrium surface vortex layer, which continues to be defined by (\ref{6.9}), generates vanishing bulk flow, hence negligible energy.

Note that such vortex layers exist also in positive temperature Euler equilibria \cite{MWC1992}, however the bounds on $|\xi|$ in that case constrain these layers to finite amplitude and finite width. Correspondingly, finite $\bar T$ suffices to generate ``blurring'' of these layers, with finite excitation of vorticity into the system interior. The bound $|\xi| < M$ used in Ref.\ \cite{TDB2014} as part of their limiting procedure for axisymmetric equilibria yields similar non-singular equilibria if one scales $\bar T \propto M^2$ \cite{foot:ptvortex}.

The $\xi$-dependent conserved integrals are given by the last line of (\ref{6.19}):
\begin{equation}
\tilde g(s) = h \bar \mu_0[p_\sigma(s,y_\mathrm{in}) + p_\sigma(s,y_\mathrm{out})]
\label{7.3}
\end{equation}
corresponding again to the entire shift in the mean of $\xi$ residing on the boundaries. Consistently, one obtains
\begin{equation}
A_D \xi_0 = -\frac{\partial F_\xi}{\partial \tilde \mu_0}
= 2 h \bar \mu_0 = \int ds \tilde g(s).
\label{7.4}
\end{equation}
with areal mean $\xi_0$ defined by (\ref{6.10}). Thus, even though the free energy correction in (\ref{7.1}) is vanishingly small, it still generates the finite contribution to the $\tilde \mu$ derivative required for consistency with the conserved integrals.

\begin{figure*}

\includegraphics[height=2.4in,viewport=20 0 440 280,clip]{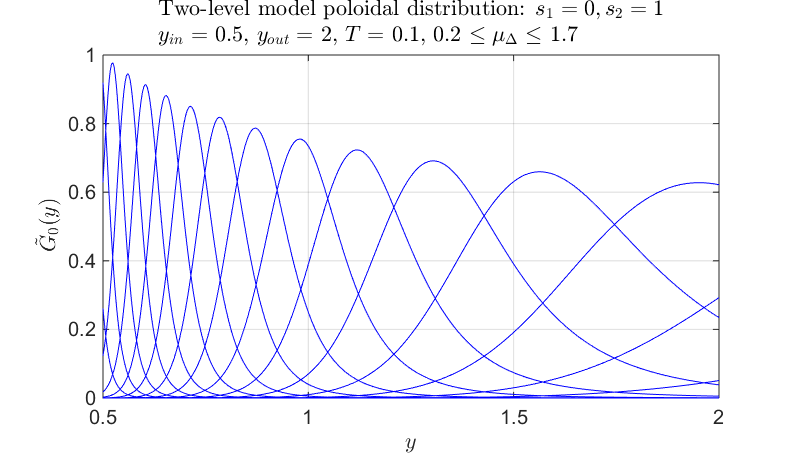}
\quad
\includegraphics[height=2.28in,viewport=0 0 390 280,clip]{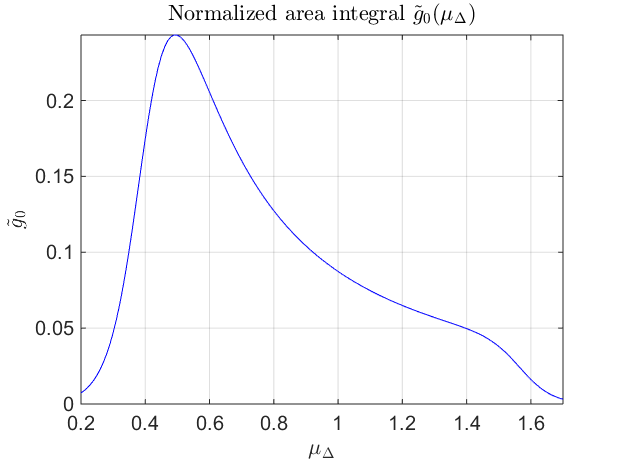}

\caption{Results for $\tilde g(s)$}

\caption{Example poloidal field equilibrium results for the same two-level model and parameters described in Fig.\ \ref{fig:2level_sigma}. \textbf{Left:} Poloidal spatial distribution $\tilde G_0(y)$ defined by (\ref{7.13}) for a range of chemical potential values $0.2 \leq \mu_\Delta \leq 1.7$ in steps of 0.1. The peak moves to left for increasing $\mu_\Delta$. \textbf{Right:} Normalized area integral $\tilde g_0(\mu_\Delta)$ defined by (\ref{7.15}). }

\label{fig:2level_xi}
\end{figure*}

\subsection{Two-level system poloidal equilibria}
\label{sec:twoleveleg}

We consider next the more interesting, non-singular problem corresponding to the two level system described in Sec.\ \ref{sec:finitelevel} [see (\ref{5.12})--(\ref{5.17})], where toroidal equilibrium quantities were also derived (Fig.\ \ref{fig:2level_sigma}). Here we extend the results to include poloidal equilibrium quantities. The four Lagrange multiplier parameters $\mu(s_1), \mu(s_2)$ and $\tilde \mu(s_1), \tilde \mu(s_2)$ reduce the general solution derived in Sec.\ \ref{sec:sigma_red} to a finite dimensional problem.

The form (\ref{5.12}) restricting $\sigma = s_1,s_2$ leads to
\begin{equation}
\tilde \mu[\sigma({\bm \rho})] = \tilde \mu_1
\chi_{\{\sigma = s_1\}}({\bm \rho})
+ \tilde \mu_2 \chi_{\{\sigma = s_2\}}({\bm \rho}),
\label{7.5}
\end{equation}
in which $\chi_A({\bm \rho})$ is the indicator function on the set $A$. Note that the scaling (\ref{6.7}) and (\ref{6.13}) still applies, but will only be imposed later.

Defining the Ising variable
\begin{equation}
\bar \sigma = \frac{2\sigma - (s_1+s_2)}{s_2 - s_1} = \pm 1,
\label{7.6}
\end{equation}
one may write
\begin{eqnarray}
\mu(\sigma) &=& \mu_0
+ \mu_\mathrm{\Delta} \bar \sigma
\nonumber \\
\tilde \mu(\sigma) &=& \tilde \mu_0
+ \tilde \mu_\mathrm{\Delta} \bar \sigma
\label{7.7}
\end{eqnarray}
in which $\mu_0,\mu_\Delta$ were defined in (\ref{5.15}), and similarly
\begin{equation}
\tilde \mu_0 = \frac{\tilde \mu_2 + \tilde \mu_1}{2},\ \
\tilde \mu_\mathrm{\Delta} = \frac{\tilde \mu_2 - \tilde \mu_1}{2}.
\label{7.8}
\end{equation}

Substituting (\ref{7.6})--(\ref{7.8}) into the discrete form (\ref{6.6}) of the $\sigma$-functional, one obtains the nearest neighbor antiferromagnetic Ising model form
\begin{equation}
\beta {\cal K}_1[\sigma] = \frac{\bar \beta}{2} \sum_{l,\hat {\bm \delta}}
J_{l,\hat {\bm \delta}} \bar \sigma_l \bar \sigma_{l+\hat {\bm \delta}}
- \bar \beta \sum_l h_l \bar \sigma_l + \bar \beta {\cal K}_0
\label{7.9}
\end{equation}
in which the $\hat {\bm \delta}$ sum in the first term runs over nearest neighbors $\pm \hat {\bf y},\pm \hat {\bf z}$, and includes dropping of boundary terms as described below (\ref{5.24}). The (antiferromagnetic) exchange, magnetic field, and additive parameters are given, respectively, by
\begin{eqnarray}
J_{l,\pm \hat {\bf y}} &=& \frac{\tilde \mu_\mathrm{\Delta}^2}{a^2},\ \
J_{l,\pm \hat {\bf z}} = \frac{\tilde \mu_\mathrm{\Delta}^2}{2 a^2 y_l}
\nonumber \\
h_l &=& \frac{s_2-s_1}{2} h_{\sigma,l} + \mu_\Delta
- \frac{\pi(s_2 - s_1)}{4y_l}
\nonumber \\
&&-\ \frac{\tilde \mu_\mathrm{\Delta}}{a^2}
\sum_m \Delta_{lm} (\tilde \mu_0 + h_{\xi,m})
\nonumber \\
{\cal K}_0 &=& \sum_l \left[\frac{\pi(s_1^2 + s_2^2)}{4y_l}
- \mu_0 - \frac{s_1+s_2}{2} h_{\sigma,l} \right.
\nonumber \\
&&\ \ \ \ \ \
\left.-\ \frac{\tilde \mu_\Delta^2}{a^2} \left(1 + \frac{1}{2y_l} \right) \right]
\label{7.10} \\
&&+\ \frac{1}{2a^2} \sum_{l,m} (\tilde \mu_0 + h_{\xi,l}) \Delta_{lm}
(\tilde \mu_0 + h_{\xi,m}).
\nonumber
\end{eqnarray}
As discussed in Sec.\ \ref{sec:sigma_red} the finite difference operations in $\Delta_{lm}$ annihilate the parameter $\tilde \mu_0$ except for boundary terms.

The Ising model (\ref{7.9}), if taken at face value, has potentially very interesting thermodynamic behavior. For example, for finite values of the exchange parameters $J_{l,\hat {\bm \delta}}$ it will undergo a magnetic transition as the temperature $\bar T$ falls below a critical value \cite{foot:aforder}. However, the physics of the fluid system lies entirely outside of this regime, with (perhaps unfortunately) the scaling (\ref{6.7}) and (\ref{6.13}) implying asymptotically vanishing exchange parameters, but still just large enough to enforce finite values of the poloidal conserved integrals.

Example equilibrium results for the toroidal field $\sigma$ were shown in Fig.\ \ref{fig:2level_sigma}, using $s_1 = 0$, $s_2 = 1$. We now extend these results to characterize equilibria including the poloidal field $\xi$. When the scaling (\ref{6.7}) is applied, all parameters vanish except $h_l = \mu_\Delta - \pi/4y_l$, agreeing with the exponential argument in the distributions $p_\sigma(s,y)$---see (\ref{5.14}).

The conserved integrals $\tilde g(s)$ may be written in the form
\begin{equation}
\tilde g(s) = A_D [p_1 \tilde \xi_1 \delta(s-s_1) + p_2 \tilde \xi_2 \delta(s - s_2)]
\label{7.11}
\end{equation}
in which $p_m$ are the fractional areas (\ref{5.17}) on which $\sigma = s_m$ (and plotted in the right panel of Fig.\ \ref{fig:2level_sigma}), and $\xi_m$ are then the mean values of $\xi$ restricted to the respective areas. From the general result (\ref{6.19}), the latter take the form
\begin{equation}
\tilde \xi_m = \frac{1}{p_m A_D} \int_D d{\bm \rho} \tilde G_m(y)
\label{7.12}
\end{equation}
with integrands
\begin{eqnarray}
\tilde G_m(y) &=& p_\sigma(s_m,y) \left(2 + \frac{1}{y} \right)
\nonumber \\
&&\times\ [\bar \mu_1 p_\sigma(s_1,y) + \bar \mu_2 p_\sigma(s_2,y) - \bar \mu_m]
\nonumber \\
&=& (-1)^{m-1} 2\bar \mu_\Delta \tilde G_0(y)
\nonumber \\
\tilde G_0(y) &\equiv& \left(2+\frac{1}{y}\right)
p_\sigma(s_2,y) [1 - p_\sigma(s_2,y)].
\label{7.13}
\end{eqnarray}
For simplicity, we have dropped the boundary term, setting $\bar \mu_0 = 0$. The integrands are equal and opposite, $\tilde G_1 = -\tilde G_2$ as required by the interior neutrality condition (\ref{6.9}), and only the difference chemical potential $\bar \mu_\Delta = (\bar \mu_2 - \bar \mu_1)/2$ enters.

The function $G_0(y)$ characterizes the spatial distribution of $\xi$-field mean values on the two toroidal (microscopically mixed) level sets $\{\sigma = s_m\}$ whose local density is given by $p_\sigma(s_m,y)$. In fact, one may identify
\begin{eqnarray}
\tilde \xi_m(y) &=& \frac{\tilde G_m(y)}{p_\sigma(s_m,y)}
\label{7.14} \\
&=& (-1)^{m-1} 2\bar \mu_\Delta
\left(2+\frac{1}{y}\right) [1 - p_\sigma(s_m,y)]
\nonumber
\end{eqnarray}
as the local mean value of $\xi$ on the respective level set at radial position $y$.

Figure \ref{fig:2level_xi} shows results for $\tilde G_0(y)$ and for the normalized area integral
\begin{eqnarray}
\tilde g_0(\mu_\Delta) &=& \frac{1}{A_D} \int_D d{\bm \rho} \tilde G_0(y)
\nonumber \\
&\Rightarrow& \tilde \xi_m(\mu_\Delta) = (-1)^{m-1}2\bar \mu_\Delta
\frac{\tilde g_0(\mu_\Delta)}{p_m(\mu_\Delta)}
\label{7.15}
\end{eqnarray}
for the same set of $\mu_\Delta$ values used in Fig.\ \ref{fig:2level_sigma}. The neutrality constraint $p_1 \tilde \xi_1 = -p_2 \tilde \xi_2$ implies that mean values $\xi_m$ must increase (decrease) as their supporting area decreases (increases). Both are linear in $\tilde \mu_\Delta$, and the one-to-one correspondence between Lagrange multiplier and conserved integral values is therefore trivial.

\section{Concluding remarks}
\label{sec:conclude}

We end by comparing the present results with the quite different axisymmetric equilibrium state predictions derived in Ref.\ \cite{TDB2014}, and suggest future numerical work that might lend insight into possible different domains of validity, depending on detailed equilibration dynamics and time scales.

\subsection{Positive temperature states}
\label{sec:posT}

In the approach taken in Ref.\ \cite{TDB2014}, in addition to the microscale $a$, a bound $|\xi| < M$ is applied, and the limit $M \to \infty$ is taken \emph{after} the limit $a \to 0$. If one limits consideration, as in the present work, to finite, positive temperatures $\bar T > 0$, no \emph{a priori} bound on $\xi$ is required, and the two limits commute. This is the domain of the full fluctuation-dominated model analyzed here (see Sec.\ \ref{sec:genmodel}). We have seen that this model provides a detailed methodology for computing candidate equilibrium states for any given values of the conserved integrals.

On the other hand, the theory proposed in \cite{TDB2014} to describe the finite $\bar T > 0$ ''low energy'' regime is based on a variational mean field approach that, unlike the exact Gaussian analysis in Sec.\ \ref{sec:statxi}, fails to account for strong fluctuations, and therefore can provide an at best approximate description. The result of that analysis is a return to the decoupled model described in Sec.\ \ref{sec:decoupledmodel}, with $\langle \xi \rangle$ trivially slaved to $\langle \sigma \rangle$. The decoupled model plays a central role in the present (exact) analysis as well, but in a very different way that leads to vanishing $\langle \xi \rangle$ and highlights instead the statistics of the large microscale fluctuations (Sec.\ \ref{subsec:gexact}, especially Sec.\ \ref{subsec:gexact}).

\subsection{Possibility of negative temperature-like intermediate states}
\label{sec:negT}

We have observed that the unbounded (Gaussian) poloidal energy $E_G[\xi]$ in (\ref{3.1}) forbids negative temperature states, $\bar T < 0$---just as in standard many particle systems with unbounded kinetic energy $\sim p^2/2m$. As a consequence, high energy initial states with, e.g., smooth large-scale (negative temperature-like) poloidal eddies, are predicted to undergo a turbulent forward energy cascade transferring much of the energy to microscale fluctuations. Specifically, the axisymmetric equations of motion (\ref{2.20}) and (\ref{2.24}) in principle provide a pathway for leakage of the flow energy into small scale (but large amplitude $\sim 1/a$) poloidal fluctuations. The analogous feature arises in the shallow water equations \cite{RVB2016,W2017}, where negative temperature states are similarly ruled out and large-scale eddy energy is expected to be transferred into small-scale surface height fluctuations (with equilibria similarly sensitive to the details of the microscale lattice geometry).

However, it is well known that such energy transfers between substantially different scales can be extremely slow, if not forbidden entirely, by the fundamental constraints of 2D flow. For example, it has long been observed, in the context of the Euler equation, that there can be strong barriers to full equilibration, with, e.g., very long-lived steady \cite{CC1996} or even fluctuating \cite{DQM2015} states preempting the statistical mechanics prediction, and depending strongly on initial condition. One may motivate this by the observation that smooth, large scale flow states, such as those described by the variational approximation, are insensitive to the microscale, and it may take time to build up the required forward cascade.

Maintaining the $|\xi| < M$ bound provides one possible route to exploring such intermediate states within the statistical equilibrium formalism. Fixed finite $M$ provides an upper bound on $E_G[\xi]$, and permits negative temperature states---the ``high energy'' regime considered in Ref.\ \cite{TDB2014}. For $\bar T < 0$ enhancement of fine-scale mixing is replaced by enhancement of smooth flows [see (\ref{6.5})]. This encourages both the $\sigma$ and $\xi$ fields to organize into high energy, macroscopic patterns (in the Coulomb analogy, like charges attract rather than repel). In particular the resulting equilibria display large-scale poloidal flow structure, including the dipolar flows and 2D vortical eddies familiar from the Euler problem. 

A consistent $M \to \infty$ limit for the mean flow may be obtained by scaling the temperature and other model parameters with $M$, in particular $\bar T^{(M)} = M^2 T^* \to \infty$ \cite{TDB2014}. It follows that the energy and mean vorticity $\langle \xi({\bm \rho}) \rangle$ remain finite even as $\langle \xi^2 \rangle \propto M^2 \to \infty$. The divergent temperature, on the other hand, ensures that the toroidal flow field $\sigma$ is completely mixed with uniform $\langle \sigma({\bm \rho}) \rangle$. Just as in the Euler case, the variational approach provides a formally exact description in this regime, and a mean field formalism may be developed in terms of the scaled variables to solve for $\langle \xi({\bm \rho}) \rangle$. For any finite $T^*$ the system is dominated by divergent fluctuations $|\xi| = O(M)$, but since we are now working in the limit $M \ll 1/a$ it is clear that the two limits do not commute in the negative temperature.

It would be extremely interesting to explore such practical equilibration questions for the axisymmetric model (as well as for the shallow water system), where buildup of microscale mixing (or surface height fluctuations), emergence of macroscale (negative temperature) coherent poloidal jet or vortex structures \cite{TDB2014}, or perhaps other features, might similarly occur on different time scales and depending on the initial state. A real fluid with finite initial $\xi$ field is unlikely to evolve toward a state accurately reproducing the large $M$ states studied in \cite{TDB2014}, even if one allows $M$ to, e.g., diverge steadily with time. However, this does not preclude similar interesting flow structures.

Proper exploration of all of these issues would probably require extended-time numerical simulation studies.

\appendix

\section{Liouville theorem}
\label{app:liouville}

Proof of the Liouville theorem for the axisymmetric system closely follows that for the Euler equation \cite{MWC1992}. A point in the infinite dimensional phase space $\Gamma$ is defined by an instantiation of the fields $\{\xi({\bm \rho}), \sigma({\bm \rho})\}_{{\bm \rho} \in D}$. The phase space gradient of a functional ${\cal F}[\xi,\sigma]$ is defined by the infinite dimensional vector of functional derivative values
\begin{equation}
\nabla_\Gamma {\cal F} = \left[\frac{\delta {\cal F}}{\delta \xi({\bm \rho})},
\frac{\delta {\cal F}}{\delta \sigma({\bm \rho})} \right]_{{\bm \rho} \in D},
\label{A1}
\end{equation}
and the phase space integral
\begin{equation}
\int_\Gamma d\Gamma = \lim_{a \to 0} \frac{1}{{\cal N}(a)}
\prod_i \int d\xi({\bm \rho}_i) \int d\sigma({\bm \rho}_i)
\label{A2}
\end{equation}
is defined by independent integration over each value of the fields at each point in $D$. Formally, it is defined here by a procedure in which the fields are first restricted to a (finite) uniform square mesh with side $a$, then the limit $a \to 0$ is taken with a suitable normalization ${\cal N}(a)$ \cite{foot:gridunits}. A probability measure $\hat \rho[\xi,\sigma]$ is a functional with unit phase space integral, and averages are defined by
\begin{equation}
\langle {\cal F} \rangle
= \int d\Gamma \hat \rho[\xi,\sigma] {\cal F}[\xi,\sigma].
\label{A3}
\end{equation}

The equations of motion are written formally in functional form $\partial_t \xi({\bm \rho}) = \dot \sigma[\xi,\sigma]({\bm \rho})$, $\partial_t \sigma({\bm \rho}) \equiv \dot \sigma[\xi,\sigma]({\bm \rho})$, and the phase space vector
\begin{equation}
{\bf W}[\xi,\sigma] = [\dot \xi({\bm \rho}), \dot \sigma({\bm \rho}) ]_{{\bm \rho} \in D}
\label{A4}
\end{equation}
defines a flow velocity field in phase space. The flow divergence is defined by
\begin{equation}
\nabla_\Gamma \cdot {\bf W} \equiv \int_D d{\bm \rho}
\left[\frac{\delta \dot \xi({\bm \rho})}{\delta \xi({\bm \rho})}
+ \frac{\delta \dot \sigma({\bm \rho})}{\delta \sigma({\bm \rho})} \right],
\label{A5}
\end{equation}
and its vanishing defines an \emph{incompressible} phase space flow. Systems with the latter property are said to obey the Liouville theorem.

A probability measure, whose time dependence $\hat \rho[\xi,\sigma](t) = \hat \rho[\xi(t),\sigma(t)]$ is defined by the evolution of the field arguments obeys the equation of motion
\begin{equation}
\partial_t \hat \rho + \nabla_\Gamma \cdot (\hat \rho {\bf W}) = 0,
\label{A6}
\end{equation}
so that the product $\hat \rho[\xi,\sigma](t) d\Gamma(t)$ is conserved, in which the (Lagrangian) volume element $d\Gamma(t)$ moves with the flow. An equilibrium measure $\hat \rho_\mathrm{eq}$ is by definition constant in time, and therefore obeys
\begin{equation}
\nabla_\Gamma \cdot (\hat \rho_\mathrm{eq} {\bf W}) = 0.
\label{A7}
\end{equation}
If phase space flows are incompressible, $\nabla_\Gamma \cdot {\bf W} = 0$, one obtains the constraint
\begin{eqnarray}
0 &=& {\bf W} \cdot \nabla_\Gamma \hat \rho_\mathrm{eq}
\nonumber \\
&=& \int_D d{\bm \rho} \left[\frac{\delta \hat \rho_\mathrm{eq}}
{\delta \xi({\bm \rho})} \dot \xi({\bm \rho})
+ \frac{\delta \hat \rho_\mathrm{eq}}
{\delta \sigma({\bm \rho})} \dot \sigma({\bm \rho}) \right]
\nonumber \\
&=& \dot {\hat \rho}_\mathrm{eq}
\label{A8}
\end{eqnarray}
which then states that $\hat \rho_\mathrm{eq}$ is a conserved integral. It follows that
\begin{equation}
\hat \rho_\mathrm{eq} = f_\mathrm{eq}(E[\xi,\sigma],
\gamma[\sigma](\cdot),\tilde \gamma[\xi,\sigma](\cdot))
\label{A9}
\end{equation}
must be some (ordinary) function of the basic set of conserved quantities, in the present case those defined in Sec.\ \ref{sec:conserve}. The choice of $f_\mathrm{eq}$ defines the statistical ensemble---see App.\ \ref{app:statmech}.

We now proceed to verify phase space incompressibility (Liouville's theorem) using the equations of motion (\ref{2.20}) and (\ref{2.24}). Since, via (\ref{2.13}) and (\ref{2.21}), the velocity field ${\bf w}$ depends only on $\xi$ one obtains
\begin{equation}
\frac{\delta \dot \sigma({\bm \rho})}{\delta \sigma({\bm \rho})}
= -{\bf w}({\bm \rho}) \cdot \nabla_{\bm \rho} \delta({\bf 0}) = 0
\label{A10}
\end{equation}
Here, one formally concludes that $\nabla_{\bm \rho} \delta({\bf 0}) = 0$ because $\delta({\bm \rho})$ is formally an even function. An alternative limiting procedure would note that the symmetric finite difference $[\xi({\bm \rho} + {\bf a}/2) - \xi({\bm \rho} - {\bf a}/2)]/a$, for arbitrary small displacement ${\bf a}$, is independent of $\xi({\bm \rho})$, hence leads to vanishing functional derivative for arbitrarily small grid cutoff $a \to 0$.

Similarly, one obtains
\begin{equation}
\frac{\delta \dot \xi({\bm \rho})}{\delta \xi({\bm \rho})}
= -{\bf w}({\bm \rho}) \cdot \nabla_{\bm \rho} \delta({\bf 0})
- {\bf w}_0({\bm \rho}) \cdot \nabla_{\bm \rho} \xi({\bm \rho})
\label{A11}
\end{equation}
in which ${\bf w}_0({\bm \rho})$ is the self-induced advection velocity for a point vortex at ${\bm \rho}$:
\begin{equation}
{\bf w}_0({\bm \rho}) = [\nabla_{\bm \rho} \times
G({\bm \rho},{\bm \rho}')]_{{\bm \rho}' = {\bm \rho}}.
\label{A12}
\end{equation}
Although the Green function $G$, obeying (\ref{2.14}) with generalized Laplacian operator defined by (\ref{2.12}), has a logarithmic singularity at ${\bm \rho}' = {\bm \rho}$, one may still derive a sensible form for ${\bf w}_0$. To see this, one separates
\begin{equation}
G({\bm \rho}, {\bm \rho}') = G_F({\bm \rho}, {\bm \rho}')
+ \Phi({\bm \rho},{\bm \rho}')
\label{A13}
\end{equation}
into free and boundary-induced parts, both symmetric in their arguments, and with $\Phi$ satisfying the generalized Laplace equation
\begin{equation}
\Delta_* \Phi = 0,
\label{A14}
\end{equation}
and chosen so that $G$ satisfies the same boundary conditions (in both arguments) discussed in Sec.\ \ref{sec:conserve}.

In the absence of a boundary, the vortex (in this case a circular vortex ring, with unit circulation, centered on the $z$-axis) self-advects vertically at constant speed \cite{Lamb,Batchelor}
\begin{equation}
v_F^z(y) = \frac{1}{4\pi r}
\left[\ln\left(\frac{8r}{a} \right) - \frac{1}{2} \right],\ \ r = \sqrt{2y}
\label{A15}
\end{equation}
which again requires a ring core radius cutoff $a$ to properly interpret, while $\Phi$ induces an additional (regular) contribution
\begin{eqnarray}
{\bf w}_\Phi({\bm \rho}) &=& [\nabla_{\bm \rho} \times
\Phi({\bm \rho},{\bm \rho}')]_{{\bm \rho}' = {\bm \rho}}
\nonumber \\
&=& \frac{1}{2} [(\nabla_{\bm \rho} + \nabla'_{\bm \rho}) \times
\Phi({\bm \rho},{\bm \rho}')]_{{\bm \rho}' = {\bm \rho}}
\nonumber \\
&=& \frac{1}{2} \nabla_{\bm \rho} \times \Phi({\bm \rho},{\bm \rho})
\label{A16}
\end{eqnarray}
in which symmetry of $\Phi$ has been used to obtain the second line.

Finally, integrating yields
\begin{eqnarray}
\int_D d{\bm \rho} \frac{\delta \dot \xi({\bm \rho})}{\delta \xi({\bm \rho})}
&=& - \int_D d{\bm \rho} [v_F^z(y) \hat {\bf z}
+ {\bf w}_\Phi({\bm \rho})] \cdot \nabla_{\bm \rho} \xi({\bm \rho})
\nonumber \\
&=& \int_D d{\bm \rho} \xi({\bm \rho})
[\partial_z v_F^z(y) + \nabla_{\bm \rho} \cdot {\bf w}_\Phi({\bm \rho})]
\nonumber \\
&&-\ \int_{\partial D} dA \xi({\bm \rho}) [v_F^z(y) \hat {\bf z}
+ {\bf w}_\Phi({\bm \rho})] \cdot \hat {\bf n}
\nonumber \\
&=& 0,
\nonumber \\
\label{A17}
\end{eqnarray}
in which incompressibility of ${\bf w}_\Phi({\bm \rho})$ follows directly from the last line of (\ref{A16}). The boundary term vanishes by virtue of the boundary conditions on $G$, which then leads to the required combination of free slip and periodic boundary conditions (described in Sec.\ \ref{sec:conserve}) on the total velocity ${\bf w}_0 = {\bf w}_\Phi + \hat {\bf z} v_F^z$.

Together, (\ref{A10}) and (\ref{A17}) establish Liouville's theorem for the axisymmetric flow system.

\section{Statistical mechanics formalism}
\label{app:statmech}

Steady state equilibrium flows are computed using phase space averages (\ref{A3}) of the flow field \cite{M1990,RS1991}, with possible forms (\ref{A9}) of the phase space equilibrium measure limited by the Liouville theorem.

The grand canonical equilibrium measure takes the exponential form \cite{MWC1992}
\begin{equation}
\hat \rho_G[\xi,\sigma] = \frac{1}{Z}
e^{-\beta {\cal K}[\xi,\sigma]}
\label{B1}
\end{equation}
in which $\beta = 1/T$ is an inverse temperature variable, and the functional ${\cal K}$ takes the form
\begin{equation}
{\cal K}[\xi,\sigma] = E[\xi,\sigma]
- \int_D d{\bm \rho} \{\mu[\sigma({\bm \rho})]
+ \xi({\bm \rho}) \tilde \mu[\sigma({\bm \rho})] \},
\label{B2}
\end{equation}
being a linear sum of all of the conserved integrals, with coefficients, or Lagrange multipliers $\{\beta,\mu(\cdot),\tilde \mu(\cdot)\}$, defining the conjugate fields. In this case there are a pair of conjugate field \emph{functions} $\mu(s), \tilde \mu(s)$ (of a single argument). The mean flow/conserved momentum parameter $v_z^0$ is suppressed here from the notation, being assumed fixed from the outset. Its only role is to introduce the constant term [first line of (\ref{3.2})].

The partition function
\begin{equation}
Z = \int d\Gamma e^{-\beta {\cal K}}
\label{B3}
\end{equation}
which simply normalizes $\hat \rho_G$, is related as usual to the thermodynamic free energy $F$ via
\begin{equation}
F[\beta,\mu(\cdot),\tilde \mu(\cdot)]
= -\frac{1}{\beta} \ln\{Z[\beta,\mu(\cdot),\tilde \mu(\cdot)]\}.
\label{B4}
\end{equation}
in which a suitable $a \to 0$ continuum limiting procedure (equivalent to the thermodynamic limit in conventional systems) is implied (see Sec.\ \ref{sec:eqmfe}). The equilibrium averages (with respect to $\hat \rho_G$) are obtained from the Free energy derivatives
\begin{eqnarray}
g(s) &\equiv& \langle \gamma[\sigma](s) \rangle
= -\frac{1}{\beta} \frac{\delta F}{\delta \mu(s)}
\nonumber \\
\tilde g(s) &\equiv& \langle \tilde \gamma[\xi,\sigma](s) \rangle
= -\frac{1}{\beta} \frac{\delta F}{\delta \tilde \mu(s)}.
\label{B5}
\end{eqnarray}
in which the conserved integrals $\gamma,\tilde \gamma$ are defined by (\ref{3.15}).

By way of comparison, the microcanonical ensemble is defined by the form
\begin{eqnarray}
\hat \rho_\mathrm{\mu}[\xi,\sigma]
&=& \frac{1}{W} \delta(\varepsilon - E[\xi,\sigma])
\label{B6} \\
&&\times\ \prod_s \delta(g(s) - \gamma[\sigma](s))
\delta(\tilde g(s) - \tilde \gamma[\xi,\sigma](s)),
\nonumber
\end{eqnarray}
in which each conserved integral is rigidly specified. The normalization $W$, which is the area of the corresponding constrained hypersurface in $\Gamma$, is related to the thermodynamic entropy via
\begin{equation}
S[\varepsilon,g(\cdot),\tilde g(\cdot)]
= \ln\{W[\varepsilon,g(\cdot),\tilde g(\cdot)]\}.
\label{B7}
\end{equation}
It is apparent that the two ensembles are related through the Laplace transform
\begin{eqnarray}
\hat \rho_G[\xi,\sigma] &=& \int d\varepsilon \int D[g] \int D[\tilde g]
\nonumber \\
&&\times\ \hat \rho_\mathrm{\mu}[\xi,\sigma;\varepsilon,g(\cdot),\tilde g(\cdot)]
\nonumber \\
&&\times\ e^{-\beta \{\varepsilon
- \int ds [\mu(s) g(s) + \tilde \mu(s) \tilde g(s)]\}}.
\label{B8}
\end{eqnarray}
in which, similar to (\ref{A2}), the functional integrals over $g$ and $\tilde g$ may be defined via a limiting procedure
\begin{equation}
\int D[g] \int D[\tilde g] = \lim_{\delta s \to 0}
\frac{1}{{\cal M}(\delta s)} \prod_l \int dg(s_l) \int d\tilde g(s_l)
\label{B9}
\end{equation}
where ${\cal M}$ is a normalization, and $s_l = l \delta s$, $l \in \mathbb{Z}$, represents a uniform gridding of the variable $s$.

Ensemble equivalence therefore reduces to the mathematical question of invertibility of this infinite dimensional Laplace transform. It is known that this property can fail in certain regions of the phase diagram \cite{BV2012}, where stable free energy minima become locally unstable saddle points. Therefore one is in general unable to access all values of the microcanonical variables through control of the conjugate field variables. However, mathematical convenience encourages one to begin with the grand canonical approach \cite{MWC1992}, which allows a more transparent exploration of the basic physics of the model, and then to subsequently investigate potential methods (e.g., some form of analytic continuation) to extend access to these regions.


\begin{thebibliography}{}
\bibitem{foot:gridunits} Strictly speaking, the $y = r^2/2$ and $z$ coordinates have different units, and a square mesh does not actually make sense. One should instead use an $a_y \times a_z$ rectangular mesh, or first rescale to dimensionless coordinates, e.g., $z/h$, $y/y_\mathrm{out}$. However, the statistical formalism used here relies only on periodicity of the mesh. So, for simplicity, and with a slight abuse of notation, we do not make this distinction here.
\bibitem{foot:AForder} Conversely, in the presence of true antiferromagnetic order, which would be expected for divergent coupling, $\gamma < 1$, and even for sufficiently large $\bar \beta$ when $\gamma = 1$, the equilibrium average $\langle \bar \mu[\sigma({\bm \rho})] \rangle$ will display finite differences between microscale neighboring sites. Hence, noting the scaling (\ref{5.22}) of the Laplacian, both sides of (\ref{6.8}) will display divergent neighboring site difference. It follows that $\xi$ displays (divergent) antiferromagnetic order whenever $\sigma$ does, scaling as $a^{\gamma-2}$. For $\gamma < 1$ this is larger than the $O(1/\sqrt{\bar \beta} a)$ fluctuation scale of $\xi$, which does not make physical sense.
\bibitem{foot:AFpsi} It follows from (\ref{6.9}) that, even in the presence of antiferromagnetic order, one has $\langle \psi_D \rangle \to 0$, though the scaled equilibrium stream function $a^{-\gamma} \langle \psi_D \rangle$ directly tracks the antiferromagnetic structure in $\langle \bar \mu[\sigma({\bm \rho})] \rangle$. For $\gamma < 1$, this produces divergent microscale circulating velocity, scaling as $a^{\gamma - 1}$, hence unphysically infinite kinetic energy.
\bibitem{foot:aforder} More accurately, the homogeneous 2D Ising antiferromagnet, with fixed $J_z(y) \to J_z^0$ (i.e., no dependence on $1/y$), undergoes a phase transition at a temperature $\bar T_c(J_z^0)$. If one now allows inhomogeneous $J_z(y) = J_z^0/y$, then at sufficiently high temperature, $\bar T > \bar T_c(J_z^0/y_\mathrm{min})$, the entire domain will be in the disordered phase, while for sufficiently low temperature, $\bar T < \bar T_c(J_z^0/y_\mathrm{max})$, it will be in the antiferromagnetically ordered phase. In between these two temperatures, $\bar T_c(y_\mathrm{max}) < \bar T < \bar T_c(y_\mathrm{min})$ there will be a phase boundary at radial coordinate $y_c(\bar T)$ satisfying $\bar T_c(y_c) = \bar T$, with the system magnetically ordered for $y < y_c(\bar T)$ and disordered for $y > y_c(\bar T)$. In the continuum limit, one is also in the adiabatic limit, and $M_\mathrm{AF}(\bar T,y) = M_\mathrm{eq}(\bar T y)$ will exactly follow the local equilibrium value of the sublattice magnetization.
\end{thebibliography}
\end{document}